\documentclass[twocolumn,twocolappendix]{aastex63}
\usepackage{amsmath}
\usepackage{svg}

\newcommand\kms{km s$^{-1}$}

\newcommand\teff{$T_{\rm eff}$}
\newcommand\logg{$\log g$}

\defcitealias{kounkel2018a}{Paper I}
\usepackage{listings}

\begin{document}
\title{Probabilistic neural network approach to determining parameters of eclipsing binaries}

\author[0000-0002-5365-1267]{Marina Kounkel}
\affil{Department of Physics and Astronomy, University of North Florida, 1 UNF Dr, Jacksonville, FL, 32224, USA}
\email{marina.kounkel@unf.edu}
\author[0009-0009-6996-4645]{Logan Sizemore}
\affil{School of Electrical Engineering and Computer Science, Oregon State University, Corvallis, OR 97331, USA}
\author{Hidemi Mitani Shen}
\affil{Department of Computer Science, Western Washington University, 516 High St., Bellingham, WA 98225, USA}
\author[0009-0003-6546-2994]{Nicholas Chandler}
\affil{Department of Computer Science, Western Washington University, 516 High St., Bellingham, WA 98225, USA}
\author{Noah Reneau}
\affil{Department of Computer Science, Western Washington University, 516 High St., Bellingham, WA 98225, USA}
\author{Ian Pourlotfali}
\affil{Department of Computer Science, Western Washington University, 516 High St., Bellingham, WA 98225, USA}
\author{Ronald L. Payton}
\affil{Department of Computer Science, Western Washington University, 516 High St., Bellingham, WA 98225, USA}
\author[0000-0002-5537-008X]{Brian Hutchinson}
\affil{Department of Computer Science, Western Washington University, 516 High St., Bellingham, WA 98225, USA}
\author[0000-0003-3410-5794]{Ilija Medan}
\affil{Department of Physics and Astronomy, Vanderbilt University, VU Station 1807, Nashville, TN 37235, USA}
\author[0000-0002-3481-9052]{Keivan Stassun}
\affil{Department of Physics and Astronomy, Vanderbilt University, VU Station 1807, Nashville, TN 37235, USA}

\begin{abstract}
Eclipsing binaries provide one of the most direct mechanisms for measuring stellar properties such as mass and radius, but historically, determining these properties has been non-trivial and computationally prohibitive. As such, only a small fraction of all eclipsing binaries for which data have been available have been fully characterized. To improve computational efficiency, we construct an uncertainty-aware neural network which can ingest phase-folded light curves in any of 50 commonly used passbands, combined with phase-folded radial velocity measurements for both primary and secondary, as well as fluxes across the spectral energy distribution to predict stellar and orbital parameters of eclipsing binaries. The model was trained to be agnostic to the presence of third light, spots (both cool and hot), and incomplete data. As the model is operating in a probabilistic framework, it is also capable of outputting uncertainties in all of the parameters. The model was trained on synthetic data, and applied to a set of $\sim$200 previously solved real eclipsing binaries to demonstrate its performance. The model is capable of determining masses and radii of eclipsing binaries with precision of $\lesssim$20\% and \teff\ with precision of $\sim$500 K in only a fraction of the time it takes the more traditional solvers. Although the resulting uncertainties are larger than what is possible to produce using more boutique analysis of individual stars, in the era of large photometric surveys, this approach allows to identify the most interesting systems, and it provides a starting point of the distributions in all of the parameters that these solvers could improve upon.
\end{abstract}


\section{Introduction}

Double-lined spectroscopic eclipsing binaries (EBs) are fundamental astrophysical laboratories that allow direct determination of properties of stars, including mass, radius, and \teff, without any model assumptions. Different tools are available for determining these parameters based on the available light curves (LCs) and radial velocities (RVs), the most widely used being PHOEBE \citep{prsa2005,conroy2020}. However, ``solving'' these systems is a non-trivial task. In order to find an optimal fit and determine reasonable uncertainties in all of the parameters, it is necessary to run a Markov chain Monte Carlo (MCMC) sampler, which can take days even when running everything on a compute cluster, and exponentially longer without a reasonable initial guess. Approaches that estimate stellar parameters through MCMC offer full predictive uncertainties that reflect observational noise and the uncertainty in the parameters of the physics-based model used in sampling. In the future, it may be possible to significantly speed up this process through the usage of machine learning side-stepping the time-consuming step of constructing individual light curves\citep{wrona2025}, but it would still require constructing a full MCMC sampler.

Fortunately, ``estimating'' these parameters can be somewhat more efficient. Periodograms allow us to identify the orbital period ($P$) of a system. An RV estimator can easily find the mass ratio ($q$) and the systemic velocity of a system ($\gamma$), and subsequently find the velocity amplitudes ($K_1$ and $K_2$), eccentricity ($e$), argument of periastron ($\omega$), projected semimajor axis ($a \sin i$), and the time of the superior conjunction ($t_0$) through fitting a simple Keplerian orbit that bears similarity to the available RV measurements.

There are also estimators that work purely with LCs. Among them, EBAI \citep{prsa2008} is a neural net that takes a binned light curve to make an estimate of a sum of relative radii ($(R_1+R_2)/a$), temperature ratio ($T_2/T_1$), inclination ($\sin i$), radial eccentricity ($e \sin \omega$), and tangential eccentricity ($e \cos \omega$). Its advantage is that it is very fast, capable of producing a guess in a fraction of a second for a given system, and it is relatively accurate in the parameters it does provide, reportedly to within 10\%. Together with the parameters for the RV estimator, it is able to produce a set of reasonable initial guesses on which PHOEBE can subsequently improve through MCMC chain.

EBAI does have a number of limitations, however.
\begin{itemize}
    \item It is designed for light curves in V band. When it was developed in 2008, it was one of the most commonly used passbands. Since then, however, more and more facilities are not using Johnson filters. In particular, many space telescopes, including Kepler, Gaia, and TESS, have their own custom passbands. Even in ground-based facilities, it is becoming less common -- e.g., ASAS-SN is now conducting its observations in g band. Since the contrast ratio between the eclipses is wavelength dependent, this creates a bias in the estimates EBAI produces.
    \item Models produced using machine learning are often deterministic, i.e., passing the data through the model would always produce the same predictions. This is the case for EBAI, and it is not capable of estimating uncertainty, and even though they are subsequently determined through MCMC, one has to cast a rather wide net around the predictions produced by EBAI to fully probe the parameter space.
    \item There are many factors that can affect light curves. Many systems are spotted, and these spots produce a noticeable signal. Some stars are affected by third light - i.e, signal from a star that is separate from EB, either as an unresolved tertiary companion, or a background star (which is not uncommon e.g., in observations with TESS due to its large pixel scale). This can degrade the quality of the predictions.
    \item It is wholly decoupled from RVs. Fundamentally, what one is trying to obtain through analyzing EBs are masses and radii, which cannot be produced by independent LC or independent RV fitting. While two sets of solutions can be interfaced with one another, this is sub-optimal in estimating the fundamental stellar properties, as two approaches can have subtle biases that skew the estimates in different directions.
\end{itemize}

Even the full MCMC treatment of EBs -- fitting both RVs and LCs simultaneously and accounting for all of the relevant parameters -- has some difficulties. While masses and radii can be reliably determined, \teff s of the individual stars are only weakly constrained, even with multi-color LCs, often straying far outside of what would be considered reasonable by comparing the resulting model to the spectral energy distribution (SED). On the other hand SED fitting has a number of degeneracies when fitting multiple stars, thus SED alone cannot be used to full constrain \teff\ either, producing a bad fit in the light curves. Thus, to measure \teff, an iterative approach is required, using PHOEBE to determine \teff\ ratio, then performing an independent SED fitting to try to constrain possible ranges of \teff\ for individual stars, then refining the PHOEBE model to be consistent with these \teff s. While it can be straightforward for some stars, in cases of spotted photospheres this can be far from trivial \citep{kounkel2024}.

In this work, we aim to rectify these issues by developing a new autonomous approach to deriving stellar and orbital parameters of EBs\footnote{\url{https://github.com/hutchresearch/EBNet/tree/main}}. In more than 15 years since EBAI was originally developed, machine learning has drastically evolved. Not only is it now feasible to have significantly more complex models that are able to handle more complex inputs, it is also possible for a model to consider probabilities in its outputs. 

\section{Data}

Neural nets require vast quantities of data on which they could be trained. These data need to be representative of the entire volume of the parameter space that the model is likely to encounter afterwards, and they need to be correctly labeled to enable the model to learn from the features presented to it. This means that it is necessary to have a large training set consisting of many light curves and RV curves, each one corresponding to a particular known set of orbital and stellar parameters. 

Although many eclipsing binaries have been solved to date, numbering in the hundreds, their number is not sufficient to reliably train a model. As such, it is necessary to instead generate synthetic time series from the stellar parameters, which is possible to do with PHOEBE.

However, a model that is trained on purely synthetic data may produce perfect predictions if evaluated on newly generated synthetic data. But if the real data have any different features in comparison to the synthetic data, then the parameters predicted from evaluating on the real data might have significant artifacts and biases. Thus, it is necessary to not only make an effort for the synthetic data to resemble any kind of real data that a model might encounter, but to also test the model performance on the real data.

In this section we describe the efforts to generate synthetic light curves that could be used for training, and to assemble the real light curves of solved EBs.

\subsection{Synthetic data}
We generate a set of synthetic EBs using PHOEBE, with randomly generated synthetic parameters

\begin{itemize}
    \item Temperature of the primary is drawn from a uniform distribution from 2500 to 12000 K. 
    \item For 1/3 of the systems, temperature of the secondary is kept independent of the primary, drawn from a similar uniform distribution. In the remaining 2/3 of the systems, temperature of the secondary is drawn from 2500 K up to temperature of the primary.
    \item Mass ratio is selected to be uniform from 0.1 to 1.
    \item Period is drawn from a log-normal distribution, with the median period of $\sim$20 days and $\sigma \log P=0.7$.
    \item Time of passage of periastron is randomly selected from phase 0 to 1.
    \item Radii for both primary and secondary are drawn from a log-normal distribution with a median radius of 2.25 $R_\odot$, and $\sigma \log P=0.45$.
    \item Inclination is drawn from a normal distribution centered at $i=90^\circ$ with a scatter of $20^\circ$.
    \item Semi-major axis is drawn from a uniform distribution up to 1000 $R_\odot$.
    \item Center of mass velocity is drawn from a normal distribution with the mean of 0 \kms, and the scatter of 50 \kms.
    \item Eccentricity is drawn from an exponential distribution with the scale of 0.2.
    \item Argument of periastron is uniformly drawn from 0 to 360$^\circ$.
    \item Spots were added to some systems. They could be placed anywhere on the photosphere, with the spot size drawn from an exponential distribution with the scale of $20^\circ$, and contrast ratio from a log normal distribution with the median of 1, and $\sigma \log t_r=0.3$. Spots were added to half of the primaries and a quarter of the secondaries.
    \item Mass, \logg, and velocity amplitude are calculated based on the above parameters. \teff\ was computed to account for the size and the contrast ratio of the spots.
    \item Roche limit was computed for both primary and the secondary. If either of the radii exceeded this limit, the system was converted to semi-detached. If both radii exceeded this limit, it was converted to contact.
\end{itemize}

It was important for the parameters to be truly independent from one another, not necessarily relying on models that would relate mass, radius, and temperature, as there are a number of cases where EBs might deviate from the models (e.g., in cases of mass transfer). We also accounted for the possibility that the primary and the secondary might be misidentified. However, this approach produced quite a lot of unphysical systems (or systems so exceptionally rare as EBs so as to not be meaningful). So we added several conditions to filter out such systems, as well as some systems that fall into the range of being unphysical due the distribution chosen above.

\begin{itemize}
    \item Masses of both stars should be between 0.05 and 10 $M_\odot$.
    \item Radii are smaller than 20 $R_\sun$
    \item $-1<$\logg$<6$.
    \item $e<1$
    \item $(R_1+R_2)/a>0.02$ to ensure that the eclipses are not too narrow to be unusable.
    \item $a \cos i < R_1+R_2$ as a first check to ensure high likelihood of systems being eclipsing.
\end{itemize}

\begin{figure*}
\epsscale{1.0}
		\gridline{\fig{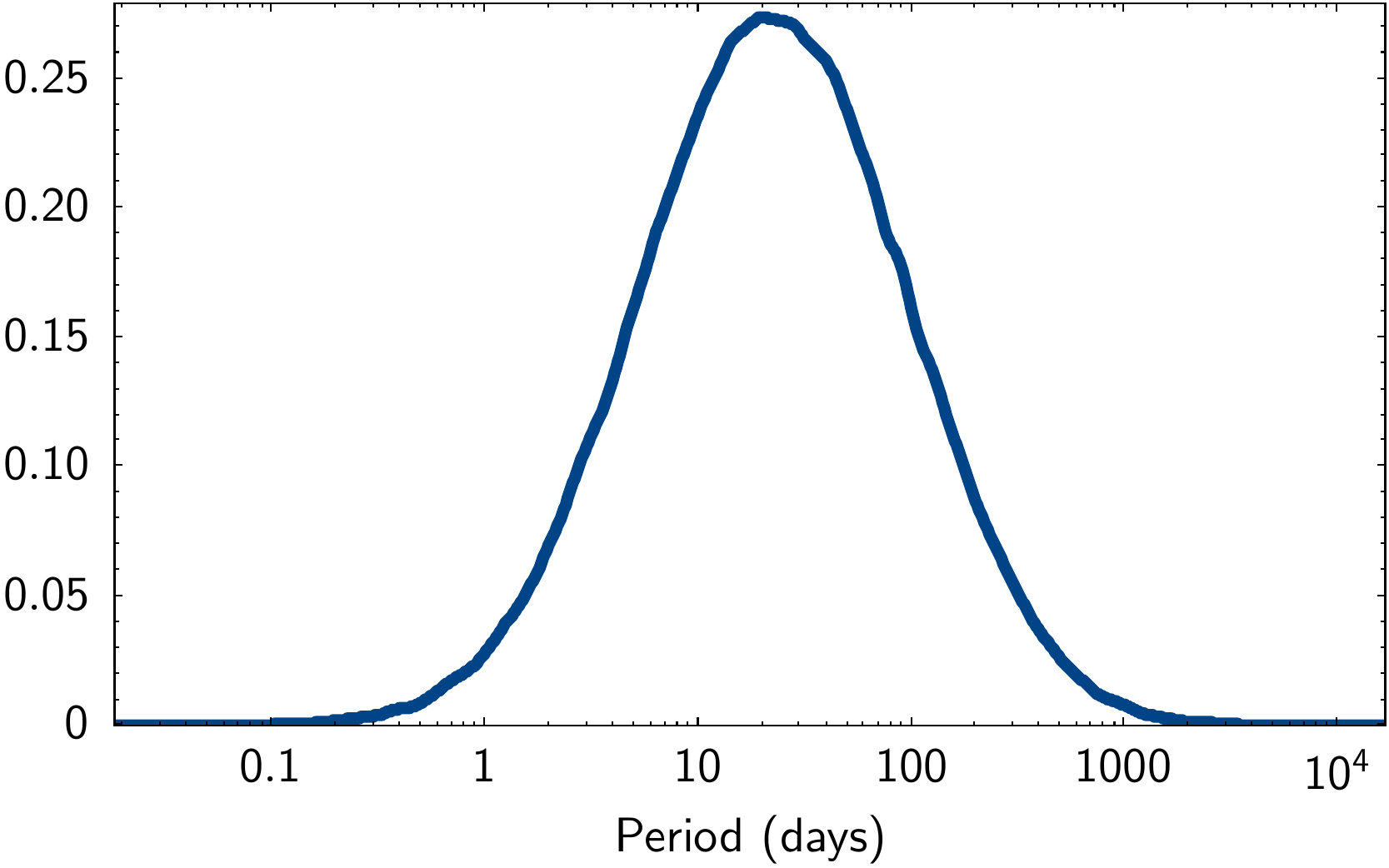}{0.33\textwidth}{}
		           \fig{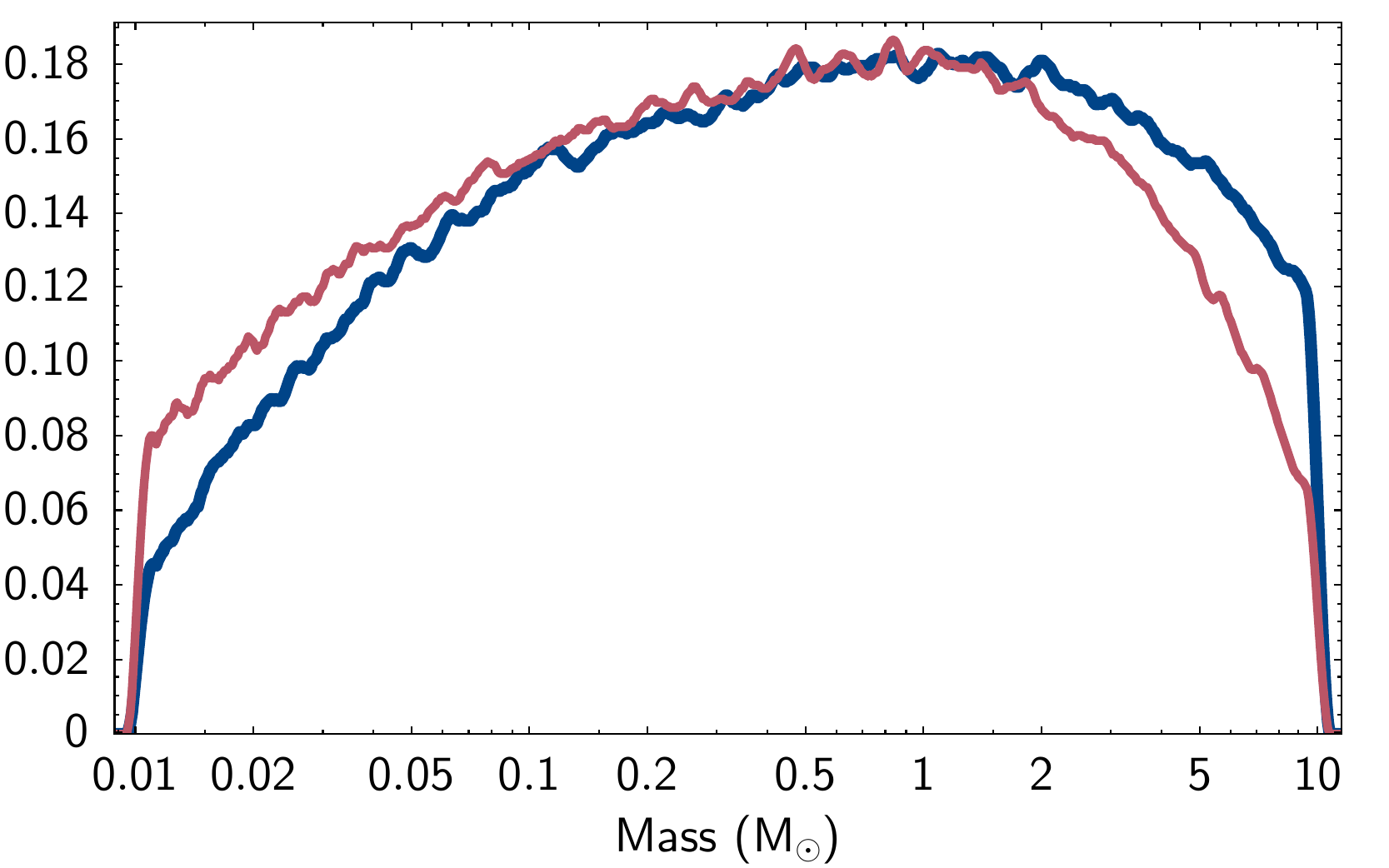}{0.33\textwidth}{}
		           \fig{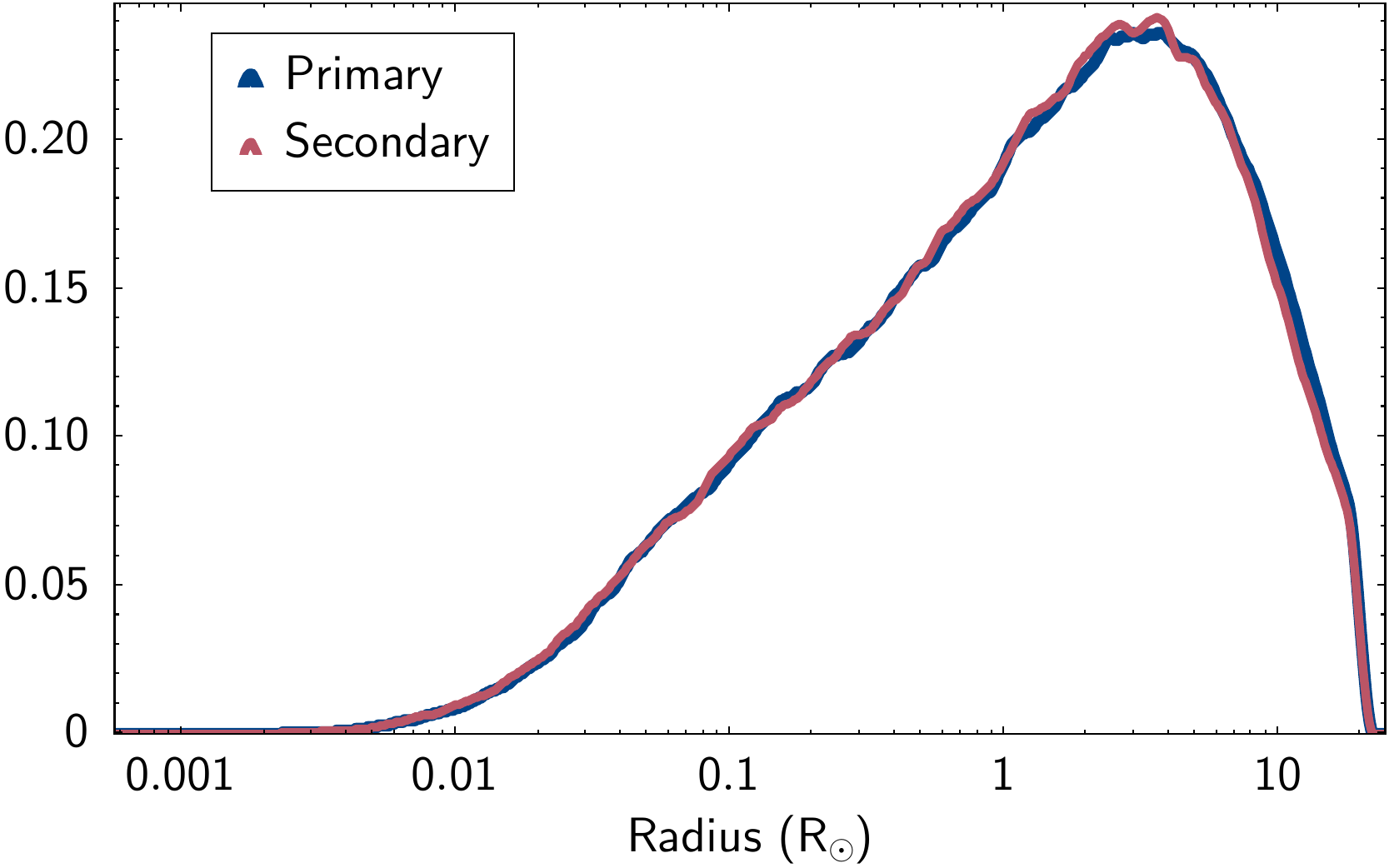}{0.33\textwidth}{}
        }\vspace{-1cm}
        \gridline{\fig{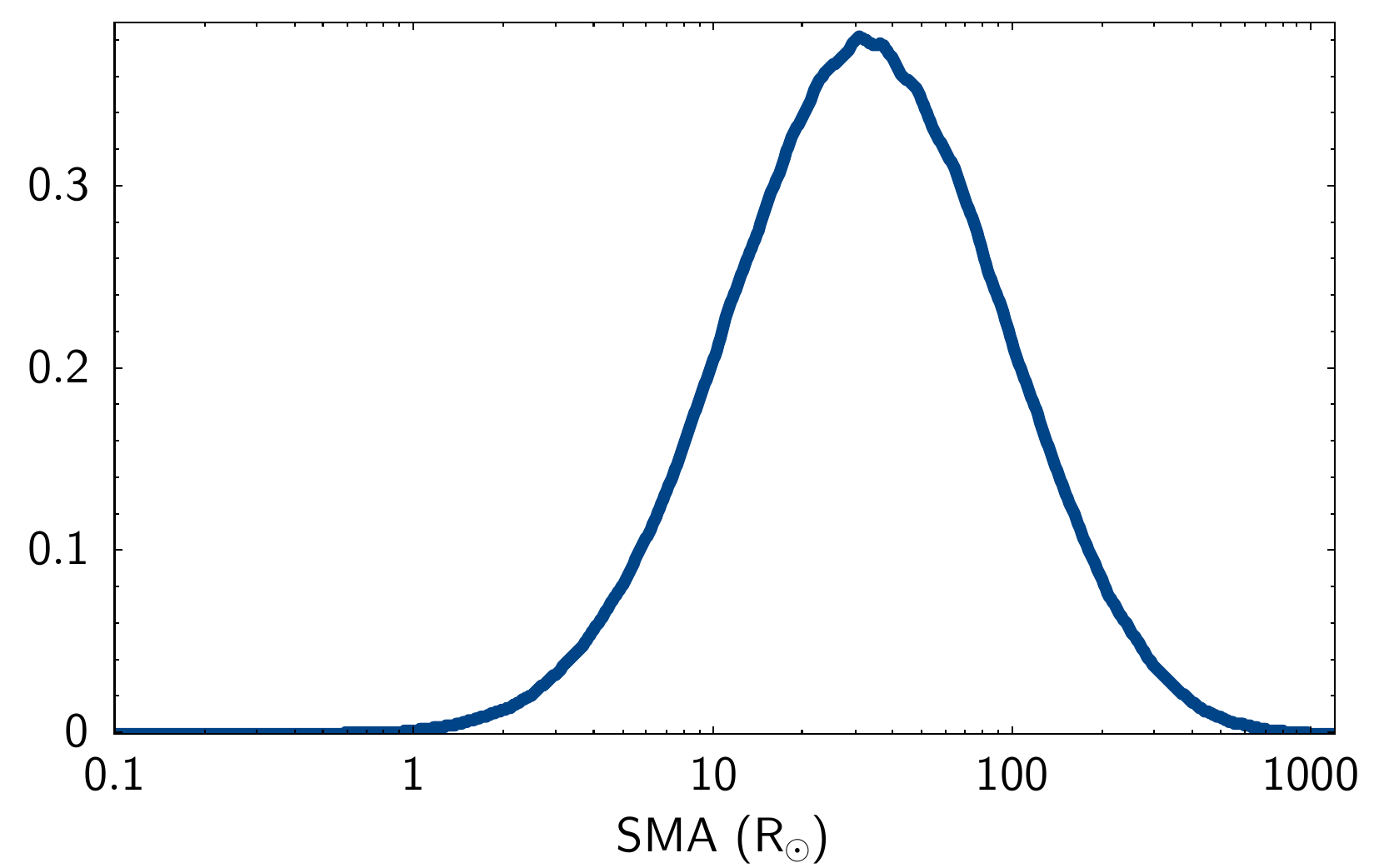}{0.33\textwidth}{}
		           \fig{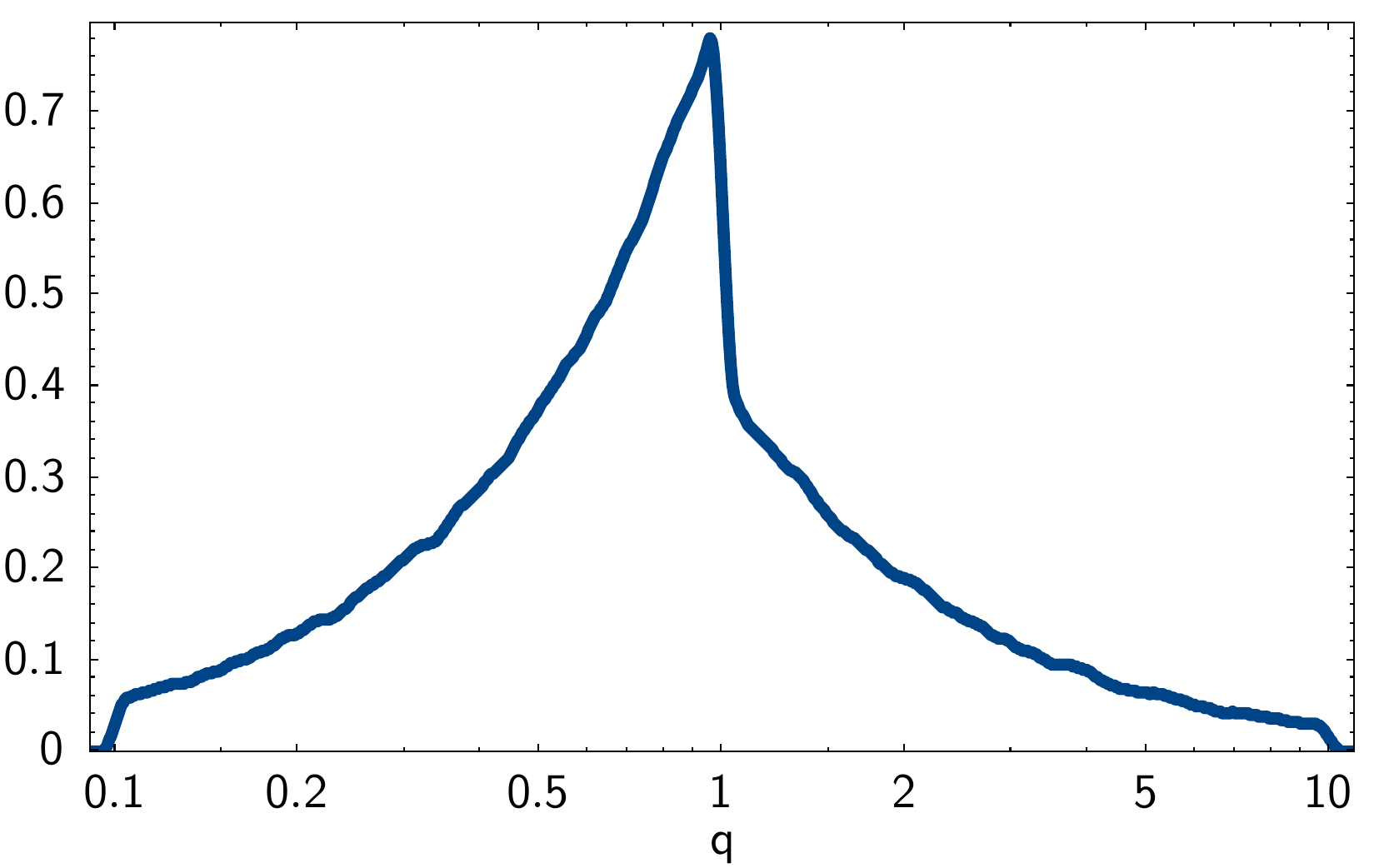}{0.33\textwidth}{}
		           \fig{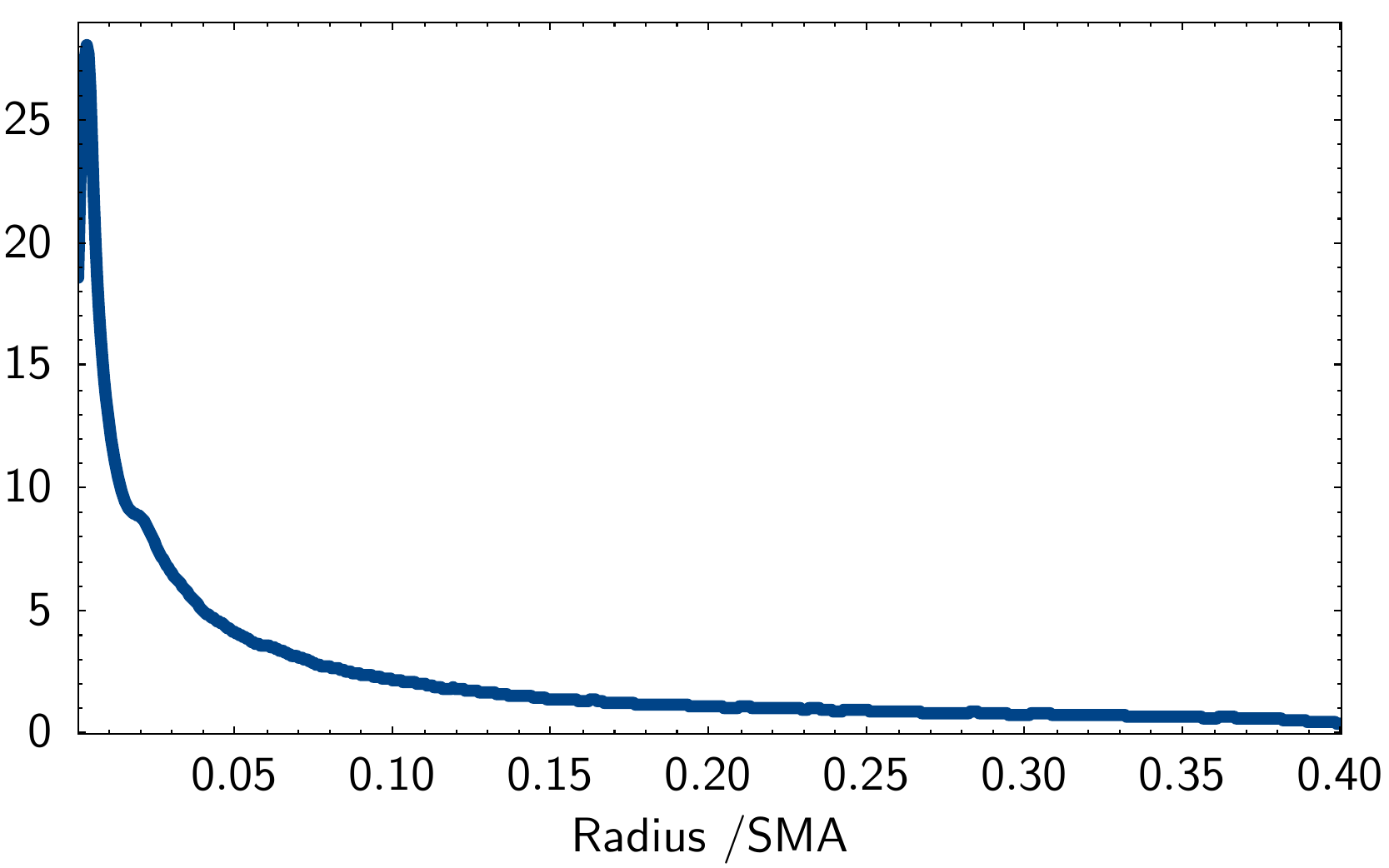}{0.33\textwidth}{}
        }\vspace{-1cm}
        \gridline{\fig{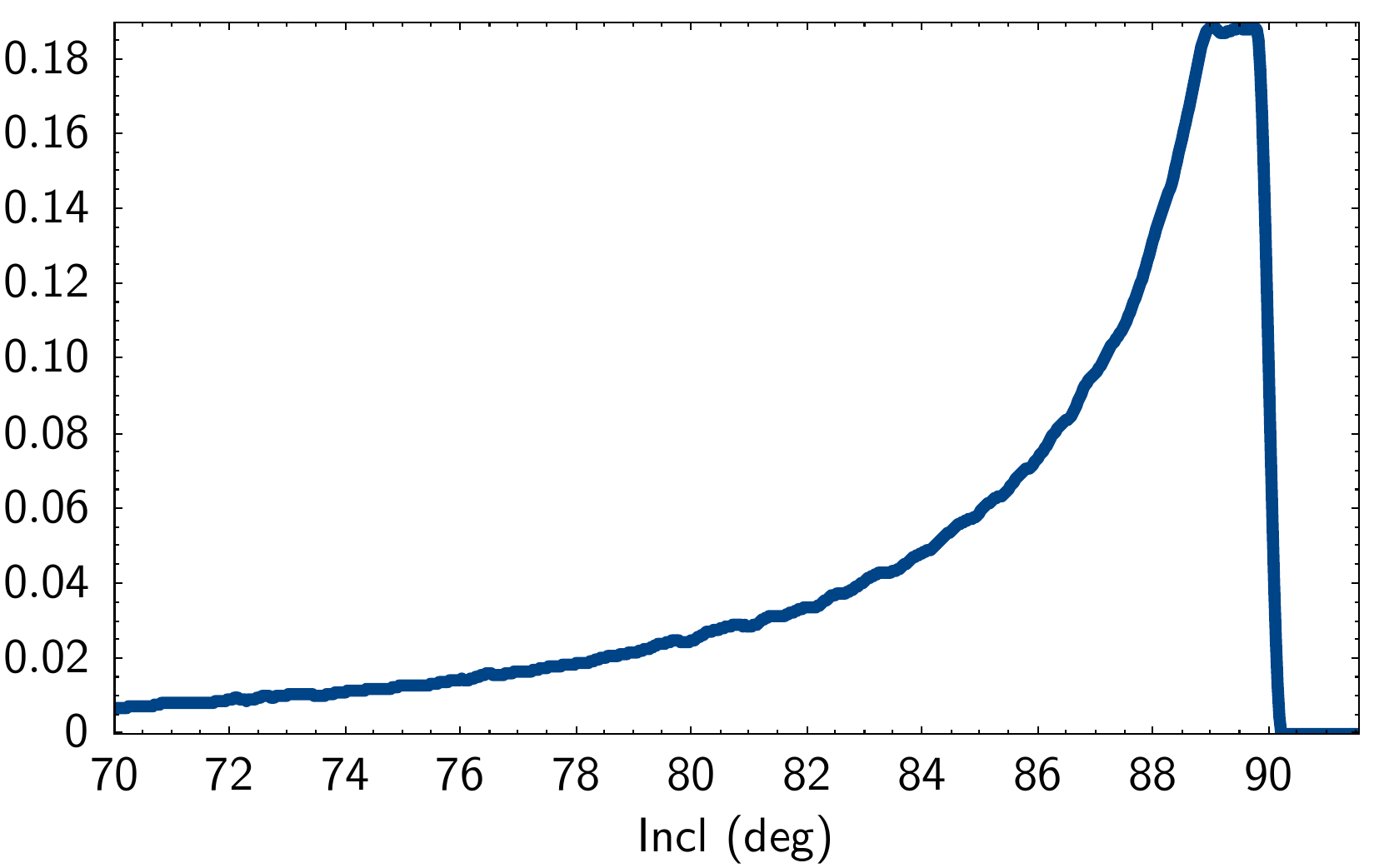}{0.33\textwidth}{}
		           \fig{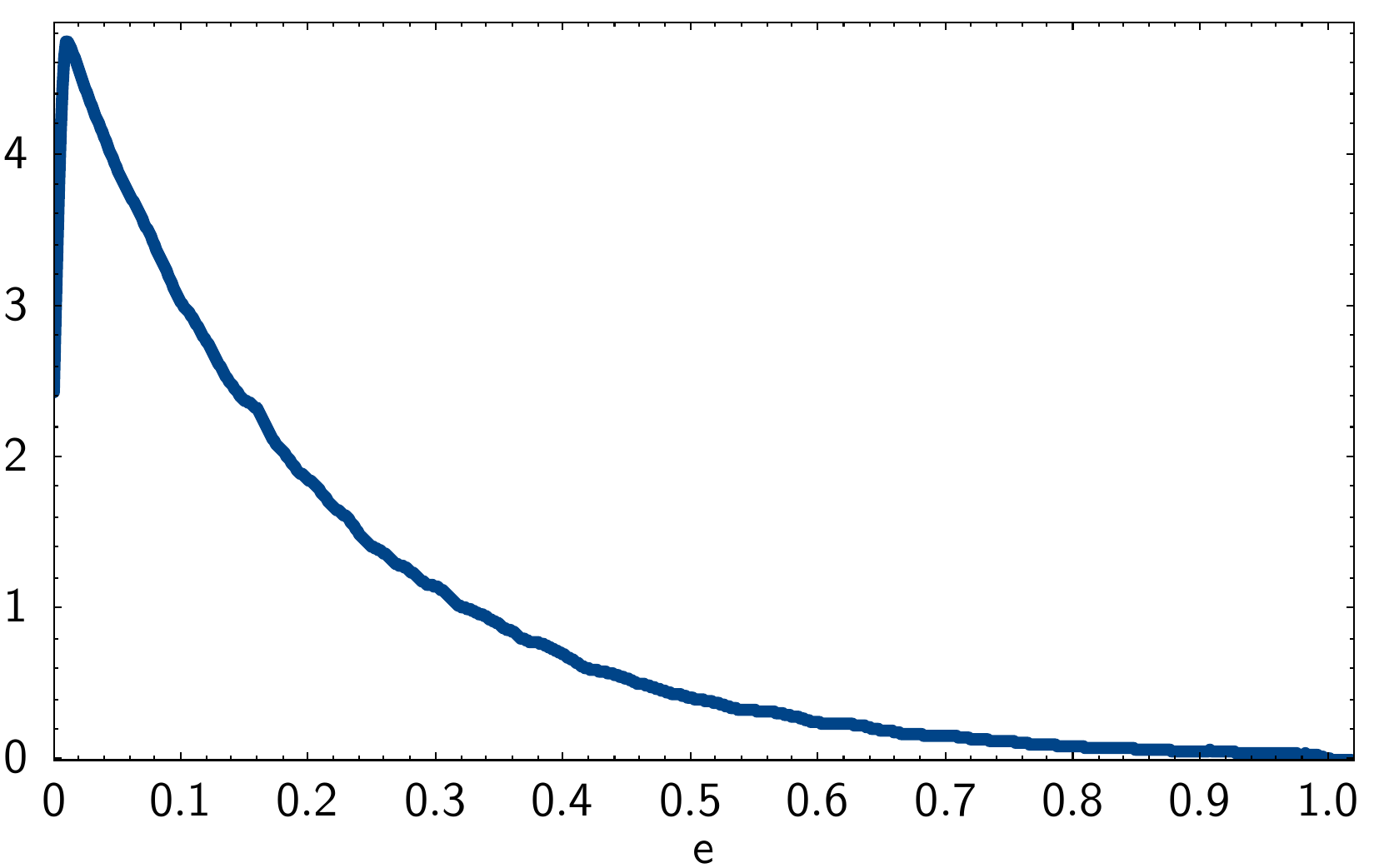}{0.33\textwidth}{}
		           \fig{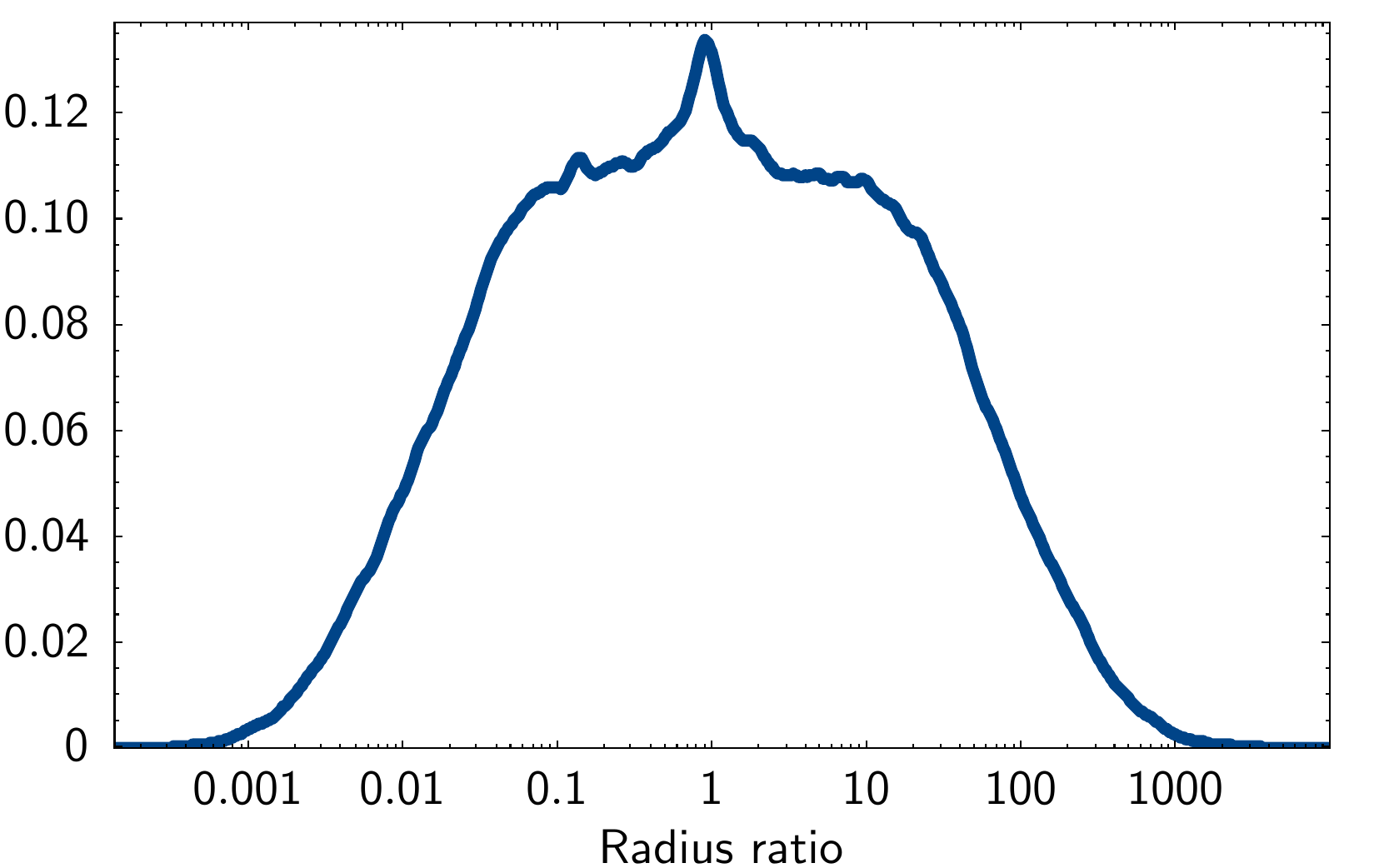}{0.33\textwidth}{}
        }\vspace{-1cm}
        \gridline{\fig{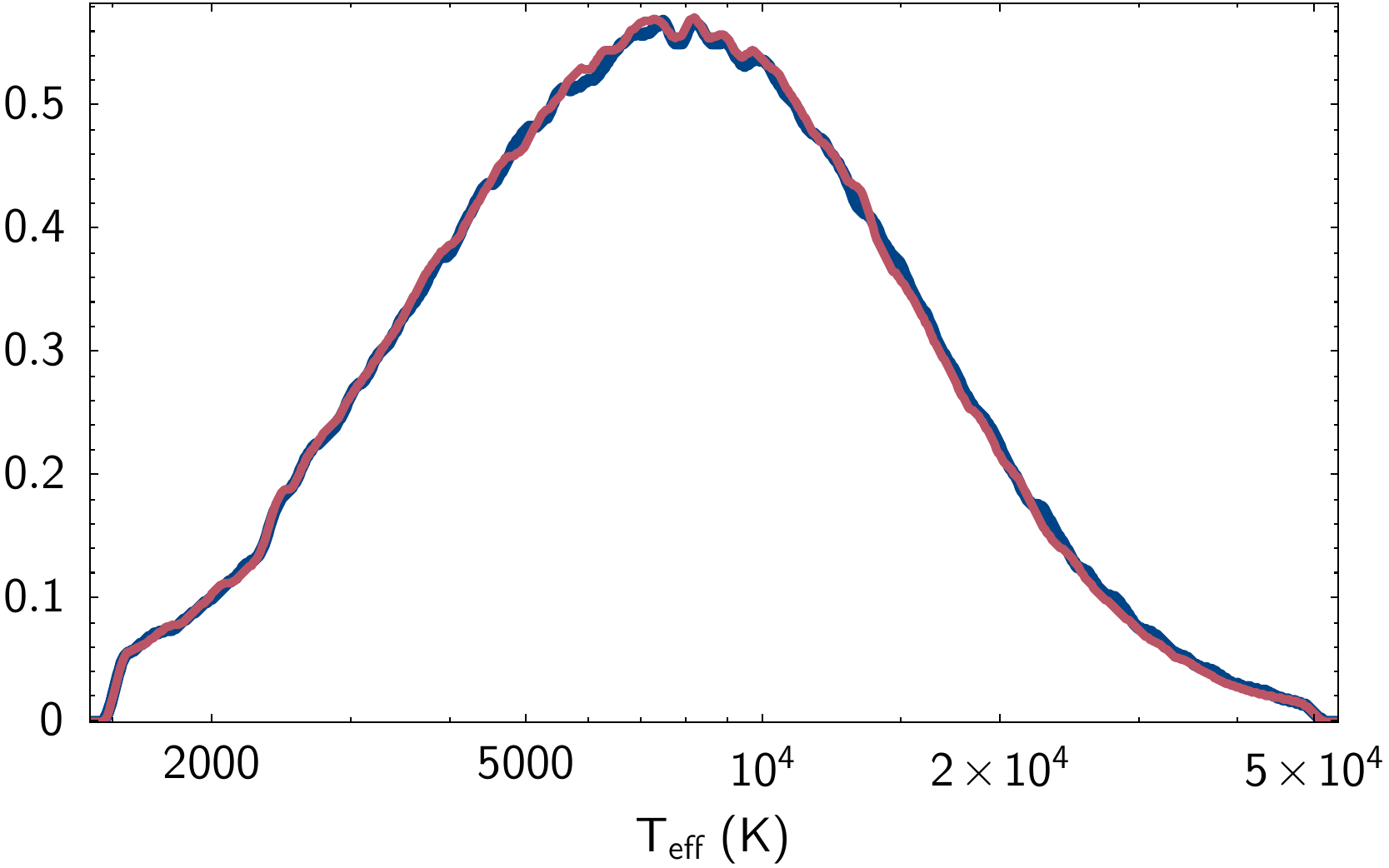}{0.33\textwidth}{}
                  \fig{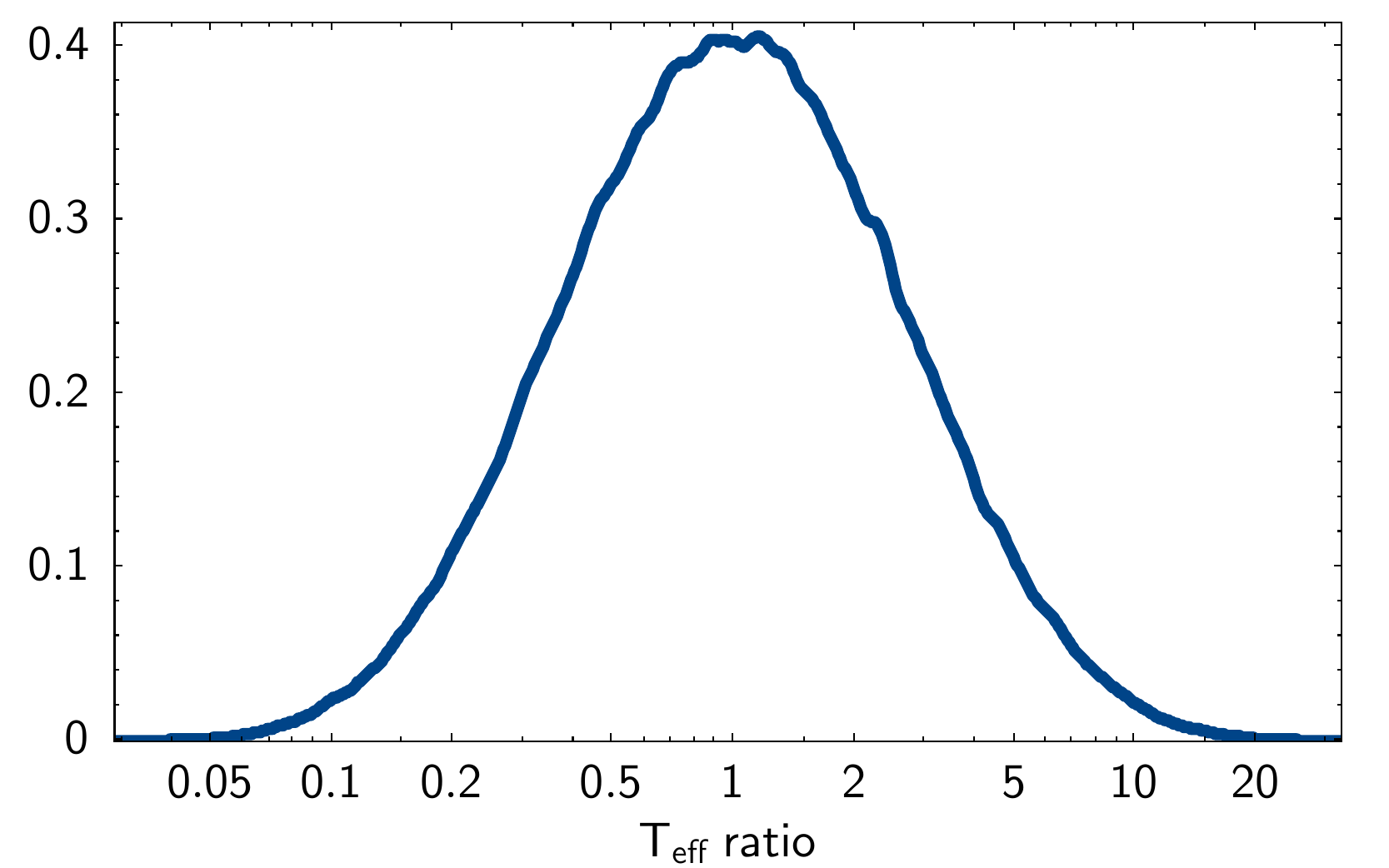}{0.33\textwidth}{}
                  \fig{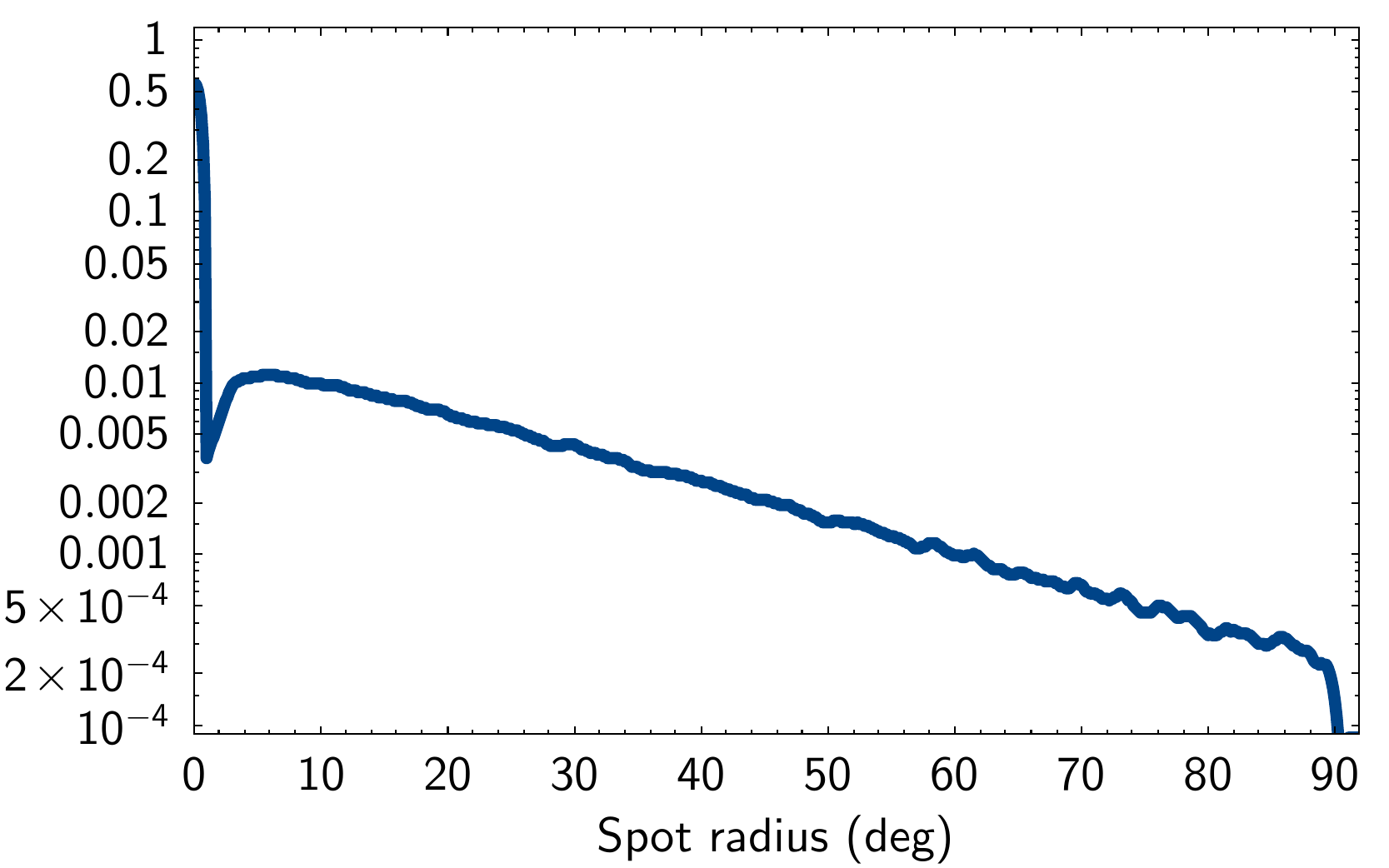}{0.33\textwidth}{}
        }\vspace{-1cm}
        \gridline{\fig{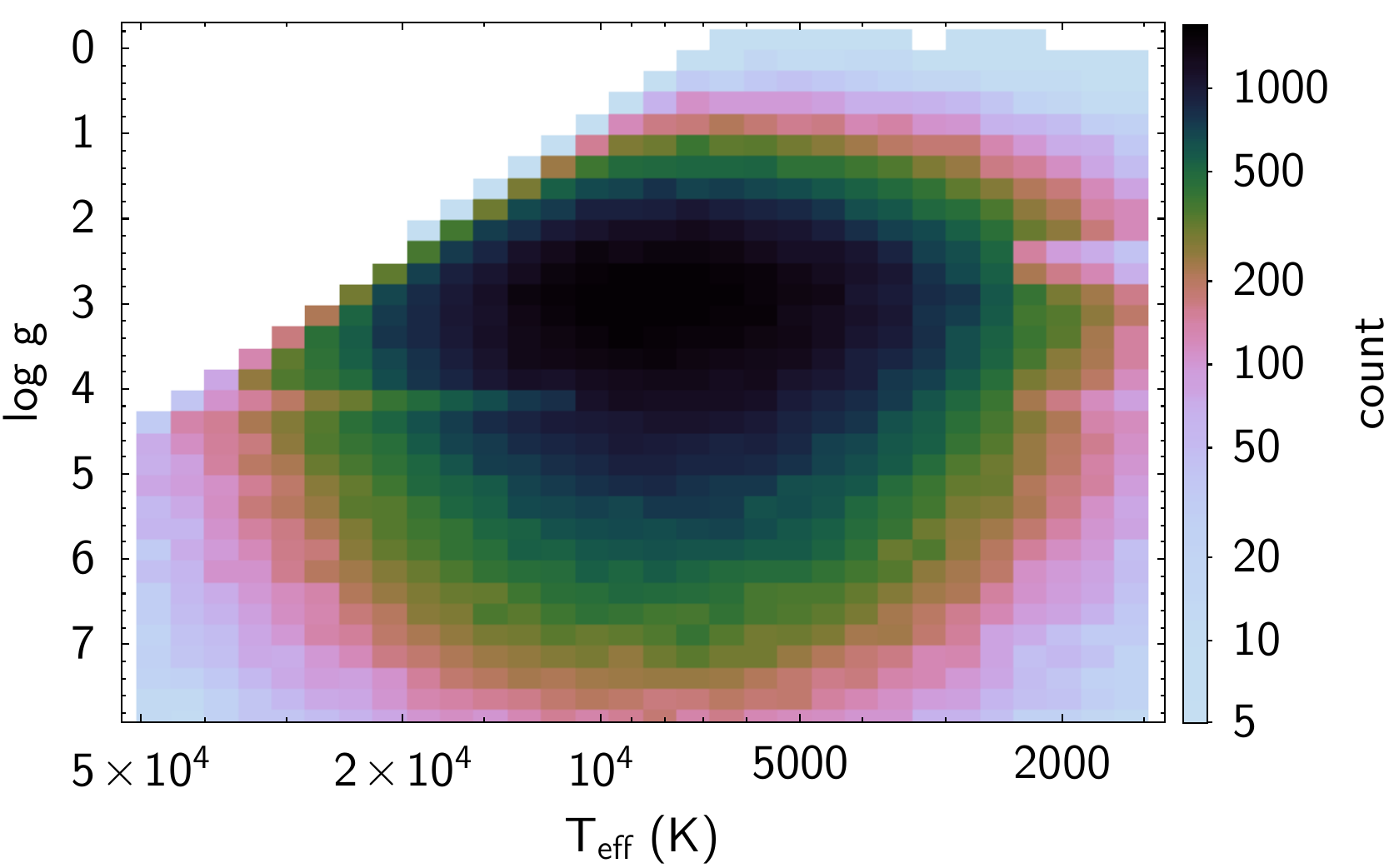}{0.33\textwidth}{}
                  \fig{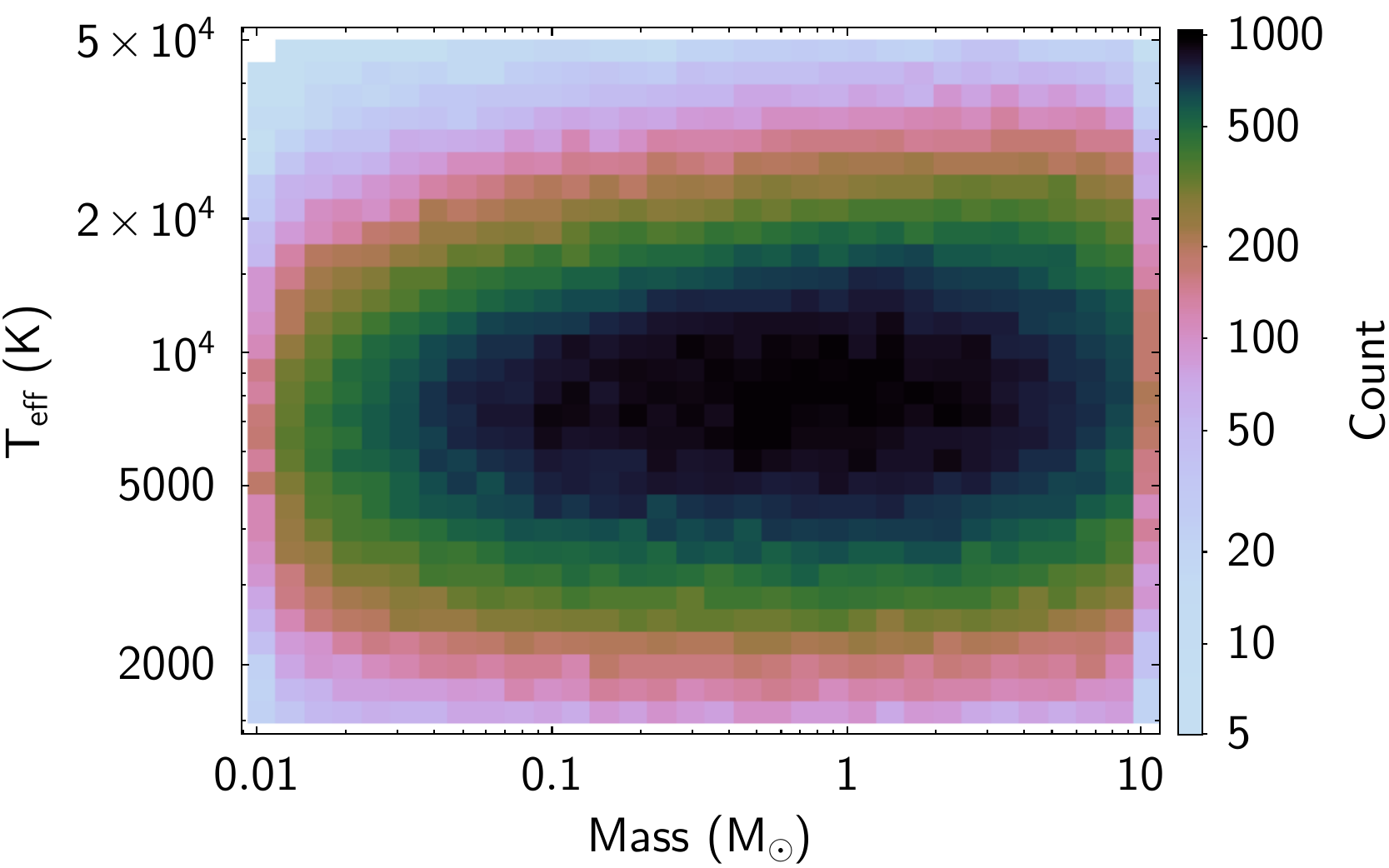}{0.33\textwidth}{}
                  \fig{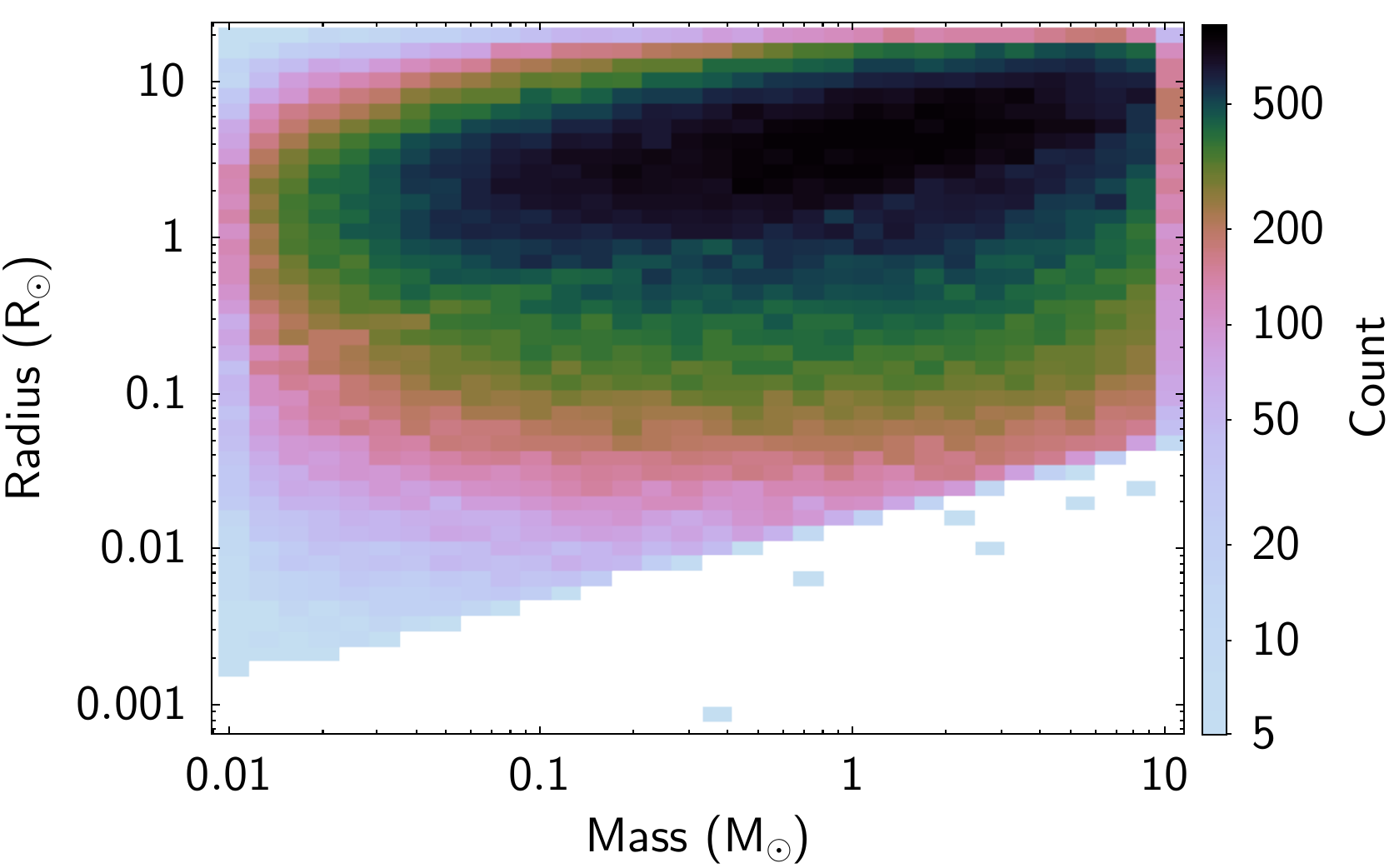}{0.33\textwidth}{}
        }\vspace{-1cm}
\caption{Distribution of the parameter space covered by the synthetic EBs.
\label{fig:synthetic}}
\end{figure*}

The resulting distribution of parameters is shown in Figure \ref{fig:synthetic}.

Using 512 equally spaced bins across the entire phase, we then generated RV curves for both primary and the secondary, as well as LCs for 50 different passbands supported by PHOEBE, including Kepler, TESS, Johnson, Cousins, 2MASS, Sloan, Gaia, PAN-STARRS, LSST, Stromgren, WASP, ZTF, and Tycho/Hipparcos filters. We used LDTK \citep{parviainen2015} to estimate the limb darkening for both of the stars in these filters given \teff\ and \logg. After generating LCs, we discarded any systems without noticeable eclipses.

Finally, we generated synthetic photometry. Using PHOENIX spectral models \citep{husser2013} we added together templates for the primary and the secondary (including their spots) with SEDFit \citep{sedfit} across 31 different bandpasses it utilizes (including Gaia, 2MASS, WISE, GALEX, SDSS, Pan-STARRS, Johnson, Cousins filters). The system is placed at an arbitrary distance, and arbitrary extinction is applied using \citet{dustextinction}.

No real EB that is observed to date likely has light curves in all 50 bandpasses - most commonly one might be able to find LCs in only one band, although some might have a handful. Only a few would have a fully sampled RV curve, and most would have around 3--20 measurements; some epochs might have detection of RV of both the primary and the secondary, and some epochs might detect only a primary. In contrast to RVs and LCs, it isn't improbable that a system would have a full complement of the fluxes across the SED, but nonetheless, some colors might still not be available for a variety of reasons.

\begin{figure*}
\epsscale{1.0}
		\gridline{\fig{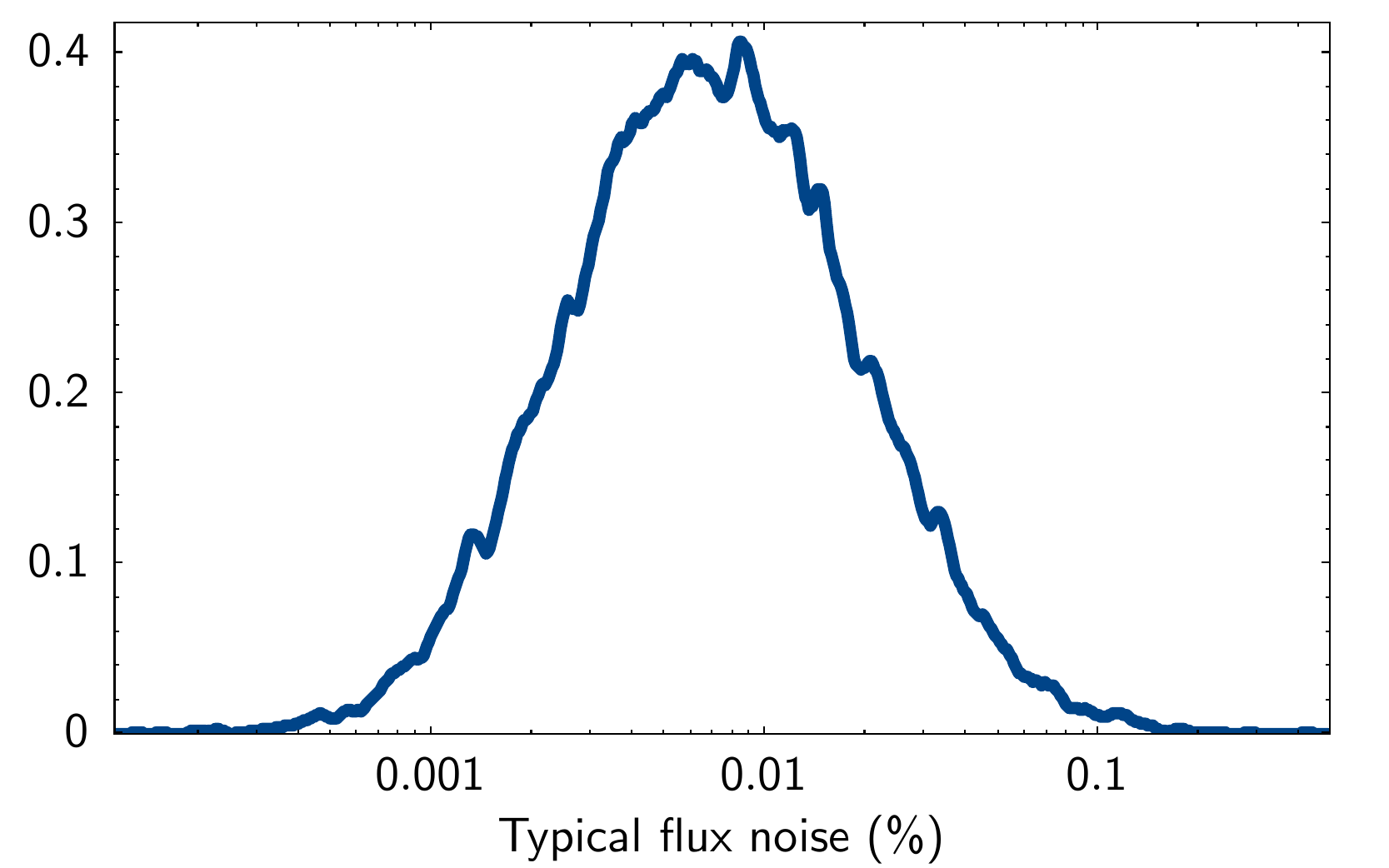}{0.33\textwidth}{}
		           \fig{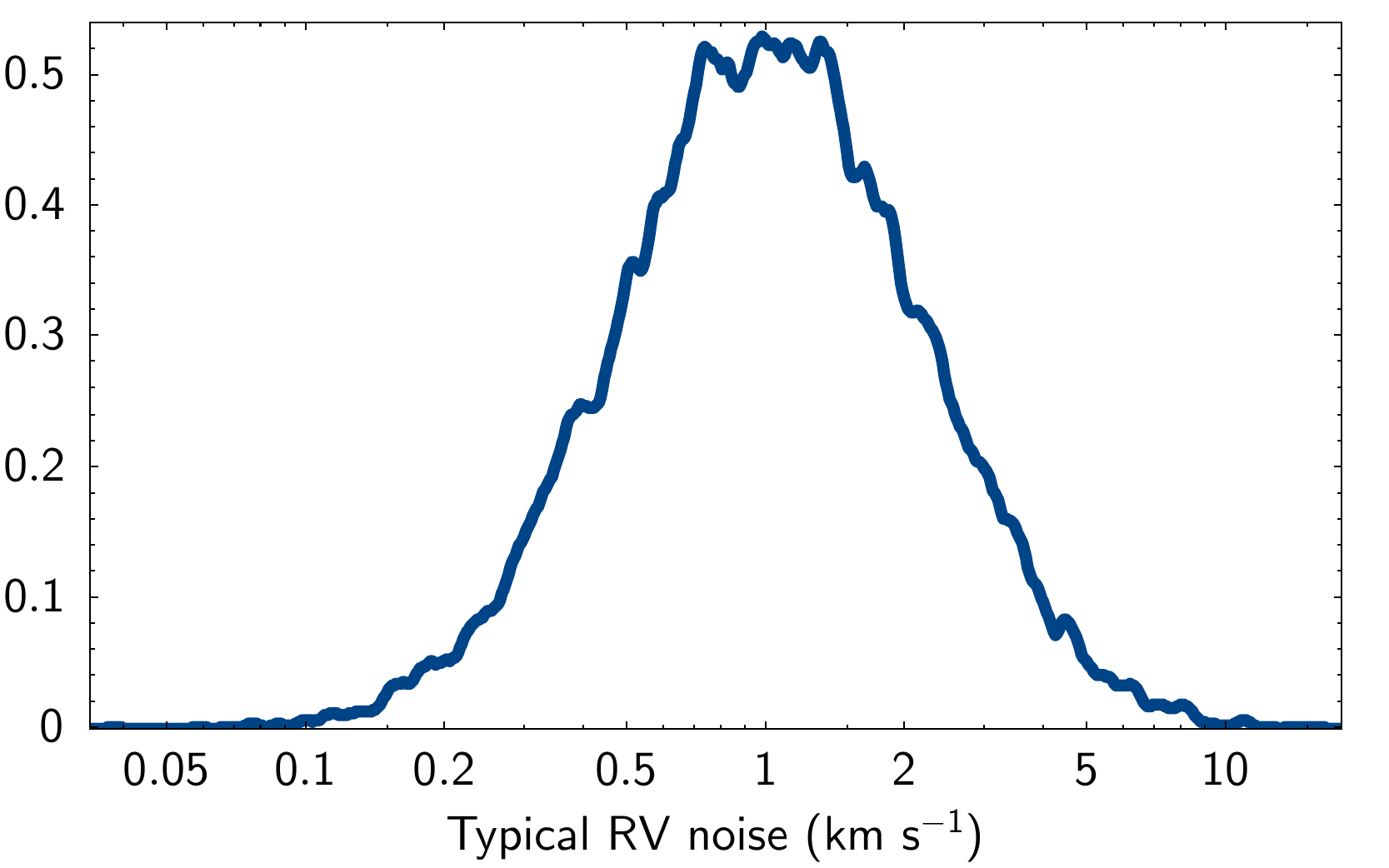}{0.33\textwidth}{}
		           \fig{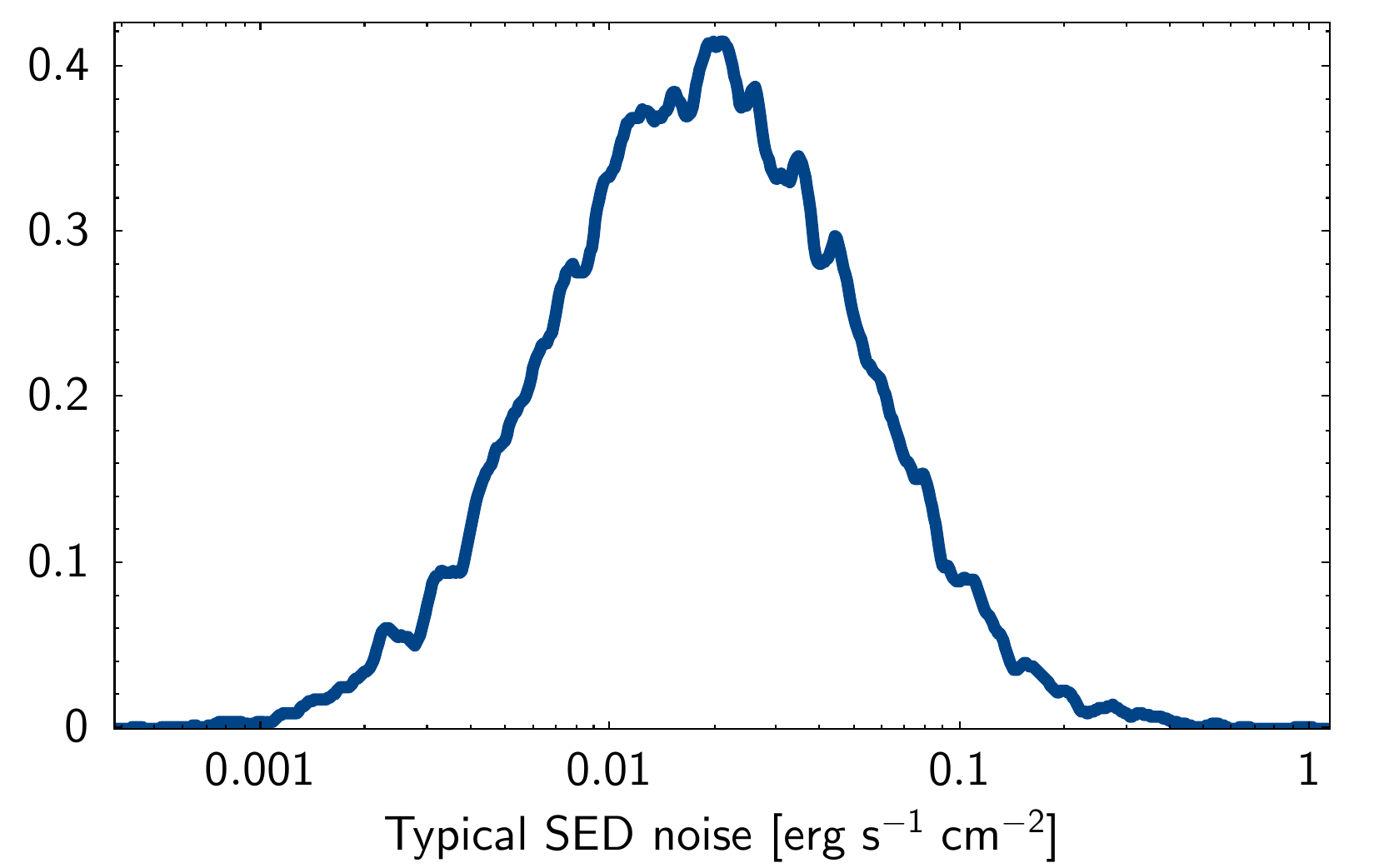}{0.33\textwidth}{}
        }\vspace{-1cm}
        \gridline{\fig{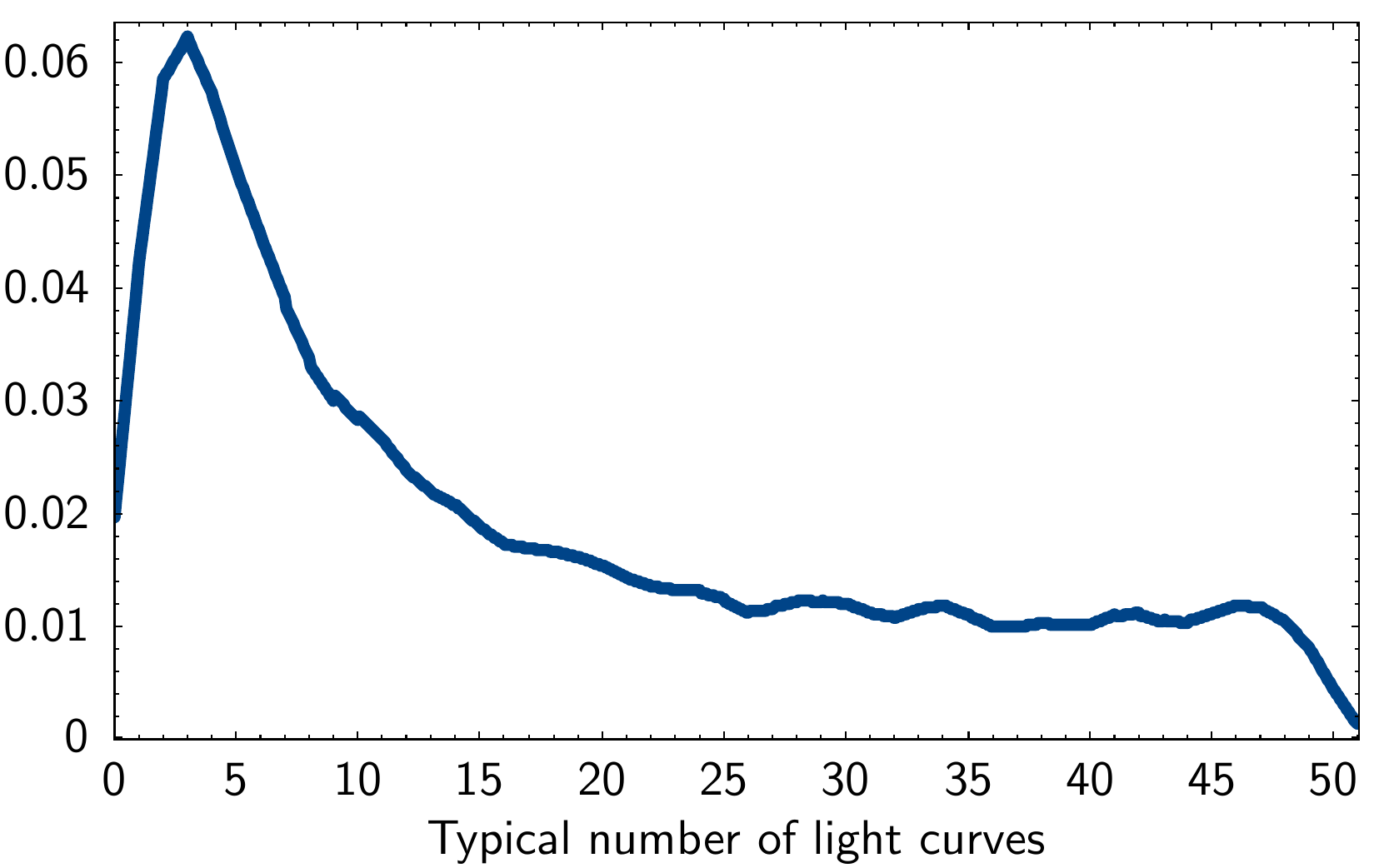}{0.33\textwidth}{}
		           \fig{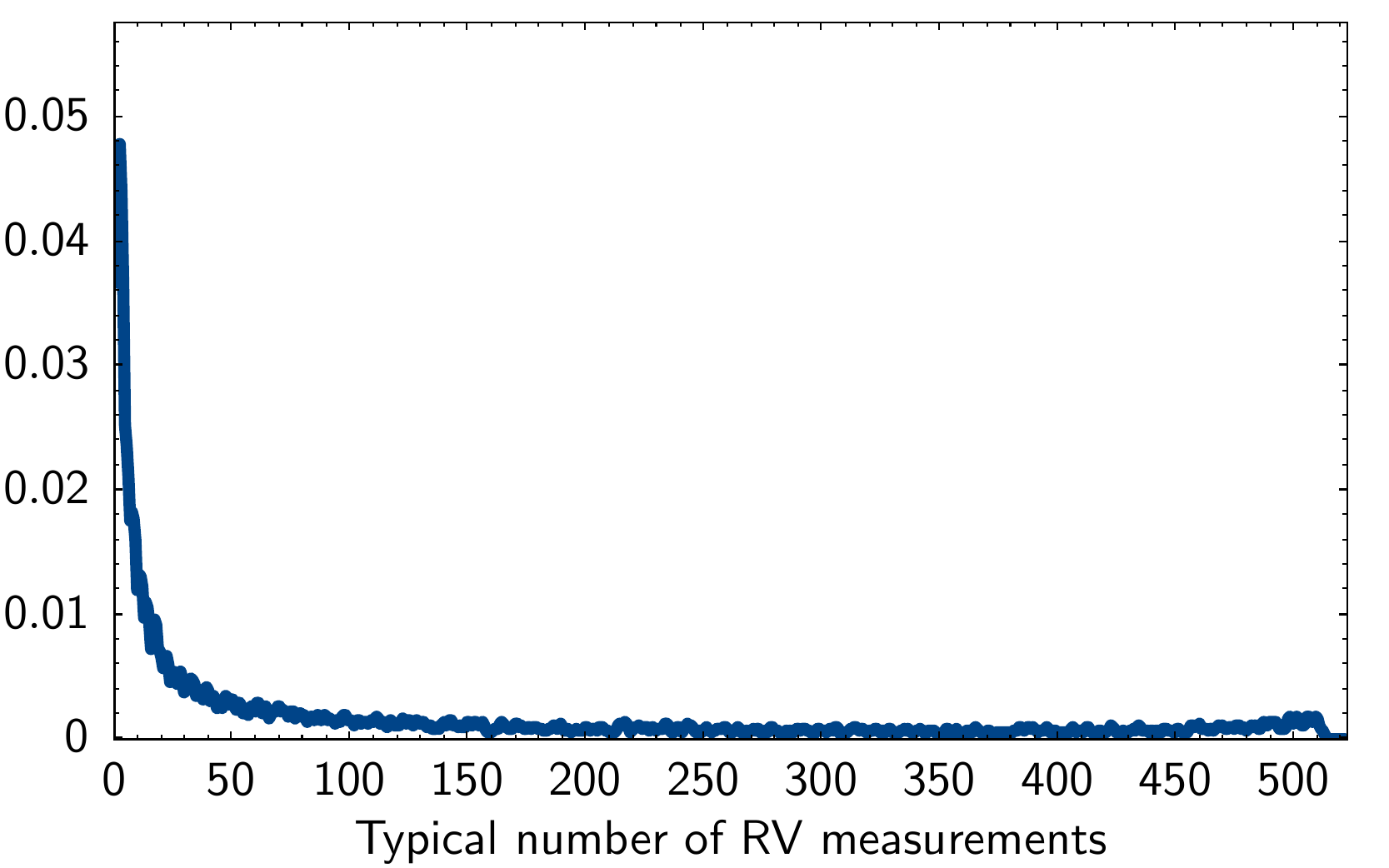}{0.33\textwidth}{}
		           \fig{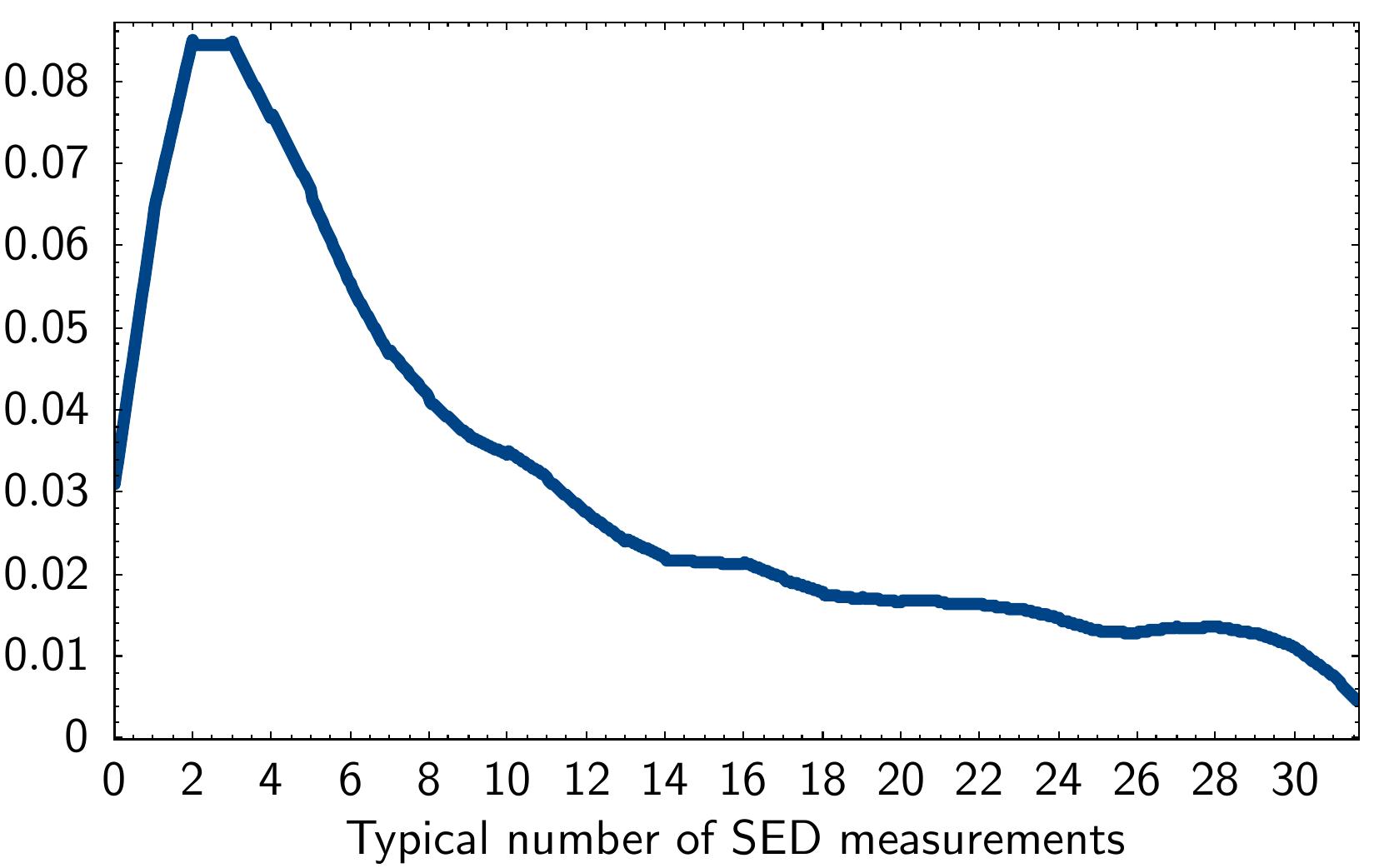}{0.33\textwidth}{}
        }\vspace{-0.5cm}
\caption{Top: typical noise distribution applied to the synthetic data. Bottom: typical size of the dataset that was not masked in training.
\label{fig:synthetic1}}
\end{figure*}

Thus, to better match these missing data, we would generate a random number of RV measurements that would be kept (sometimes both the primary and the secondary, sometimes either primary or either secondary RVs). Similarly, we would select a random number of integrated fluxes to keep as metadata, and we would select a random number of light curves for a given source to be used in training, with each bandpass having an equal probability of being chosen. If data points were dropped out, they were set to zero. However, if the light curve was selected, it was expected to be fully sampled with all 512 points. The typical distribution of the data points used in training is shown in Figure \ref{fig:synthetic1}. There was always a possibility that no RVs, no light curves, and/or no metadata would be supplied by the data loader, to prevent the model from exclusively relying on any single dataset.

Light curves were also adjusted to add a random fraction of third light to them by adding a constant to the light curve continuum, and then normalizing by the median. This resulted in some sources having significantly shallower eclipses, to replicate contamination from a star along the same line of sight.

Finally, we generated a random noise profile for each source through first generating a magnitude of a typical uncertainty for a particular dataset, and then generating a Gaussian distribution for every point in the dataset, and multiplying it by this typical error. This enabled the model to generate different types of sources with both high and low signal to noise ratio, without biasing the model to just clean and noiseless light curves or ideal RVs (Figure \ref{fig:synthetic1}).

In total we have generated $\sim300,000$ sources for training, and each source had five different realizations of passbands and the noise profile. This number of synthetic sources was sufficiently large such that increasing it further did not produce any improvements in the quality of the predictions, at least given the noise profile of the training data.

In addition to the light curves, RV curves, and integrated fluxes, we also stored additional metadata, which included the orbital period, as it could not be inferred from a phase-folded light curve, but it should be known for all of the systems, and it is needed for determining parameters such as the semimajor axis. Additionally we included a parameter for some systems of the parallax to the source (that was used for transforming the integrated fluxes), and the uncertainty that was applied to the parallax. This was done to potentially help it strengthen the determination of radii from the SED, however, both of these were treated as optional, and they were masked out in a third of all of the systems.

\subsection{Real data}\label{sec:real}

To verify the performance of the model, we have assembled a catalog of 220 solved EBs. Approximately half of the systems were the sources from \citep{eker2014}, which provided their stellar and orbital parameters, as well as radial velocities. TESS light curves were produced from the SPOC pipeline \citep{jenkins2016}.

The remaining sources were cataloged in DEBcat \citep[and references therein]{southworth2015}. The selected sources were required to be binaries, without a known third companion. All of the systems in this set have stellar and orbital parameters that are encompassed by the range of the parameters of the synthetic data.

This resulting catalog of real EBs includes many of the sources that have been initially analyzed using Kepler light curves, or using ground based photometry. In addition to the stellar and orbital parameters, we have downloaded the reported RVs from the respective publications that have performed the initial orbital fit, as well as the light curves that were used in that fit, if the light curves were tabulated in the source papers. They were then supplemented with the light curves from TESS or Kepler generated using lightkurve \citep{lightkurve}. In total, 192 sources had TESS light curves, and 33 had Kepler data. Additionally, 3 lightcurves were in Johnson B band, 20 in V band, 12 were in R band, 15 in I band, 1 in J band, 1 in K band. Other filters represented in the data include 8 sources with Stromgren u, 9 with Stromgren v, 10 with Stromgren b \& y bands, 2 sources with SDSS g' band, 1 source with SDSS r band, 1 source with SDSS r' band, and 1 SWASP lightcurve. 

\section{Model}

\begin{figure*}[t]
  \centering
  \begin{minipage}[t]{0.48\textwidth}
    \vspace{0pt}
    \centering
    Model 1\\[3pt]
    \includegraphics[width=\linewidth]{tf_model.pdf}
  \end{minipage}
  \hfill
  \begin{minipage}[t]{0.48\textwidth}
    \vspace{0pt}
    \centering
    Model 2\\[3pt]
    \includegraphics[width=\linewidth]{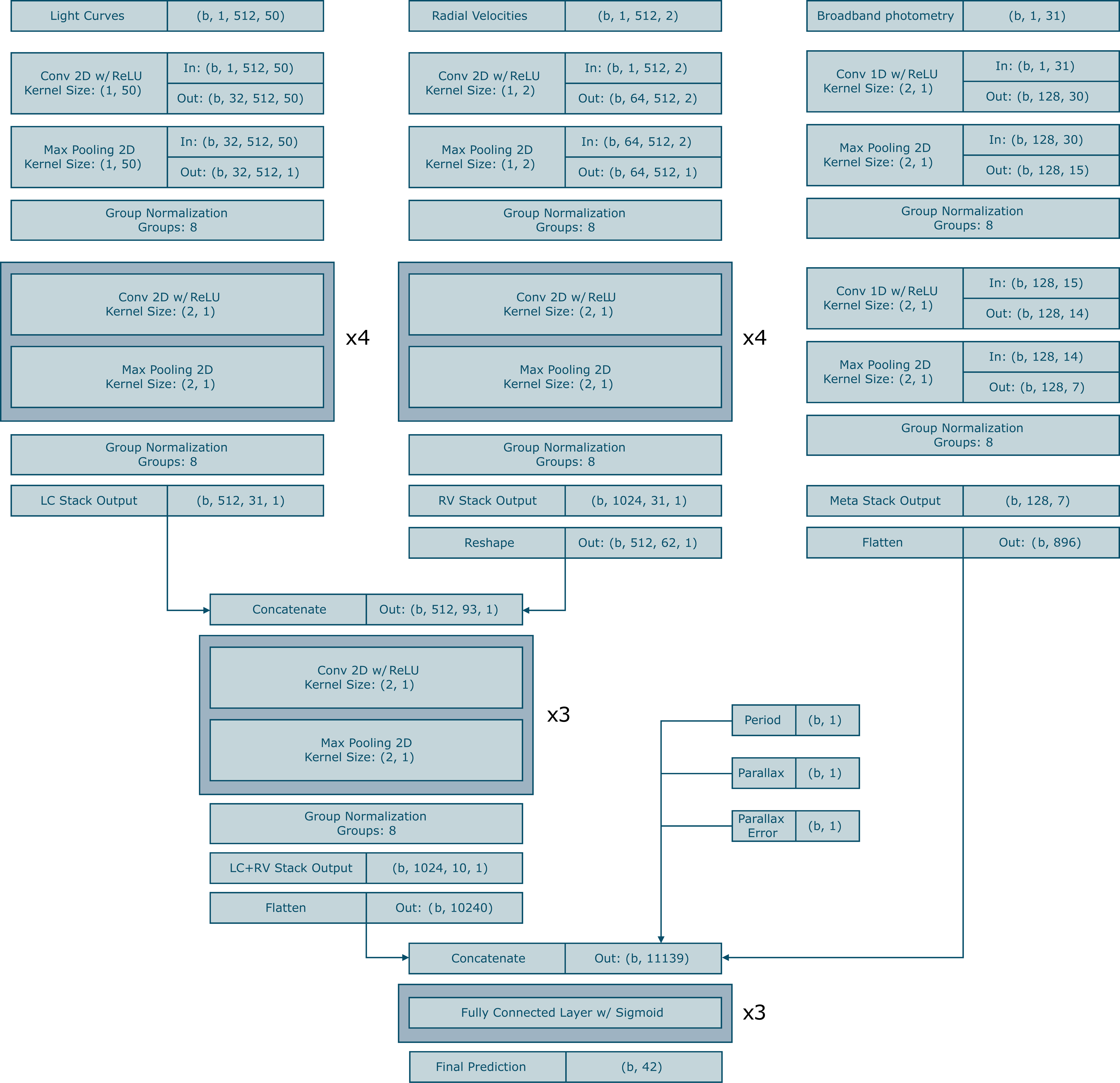}
  \end{minipage}
  \caption{Model diagrams showing the architectures of both models. The architecture of the models, as well as the hyperparameters associated with it were a result of various sweeps done to improve the overall performance.\label{fig:model}}
\end{figure*}

There were two parallel efforts to construct the neural network that is capable of processing the data on the eclipsing binaries, one of which was constructed in TensorFlow, the other in PyTorch. The models were trained on the same data, and with a similar architecture, but different hyperparameters. These hyperparameters, including the exact layer architecture, number of neurons in each layer, the choice of the activation function, etc, were determined through a series of sweeps across various models with the goal of improving the overall performance.

There were two parallel efforts to perform these sweeps, one of them was done in TensorFlow, and the other one in Pytorch. These efforts began with a similar architecture that was refined in these sweeps.

Although it was an initial goal for these efforts to culminate in a single most optimal model, instead we arrived at two separate models, one trained in each of these code-bases. While both of these models can produce a roughly comparable performance, they ended up specializing more in a particular set of parameters, and thus there was no definitive winner between them. In this section we describe their respective architectures, and in Section \ref{sec:comparison} we compare the differences in their performance.

\subsection{Model 1 (Tensorflow)}
As there are multiple types of data, each one is processed by the model separately via convolution layers, before being combined by the fully connected layers.

Since the array with the light curves consisted of primarily rows that were zeroed-out, with only one or two rows containing valid data, the initial step would aggressively convolve and pool together all 50 light curves to create 64 channels with a single row with the width of 512 (corresponding to the initial number of points in the light curve). Such aggressive downsampling was done to solely force the model to learn the position of all of the passbands that light curves may come in without necessarily relying on any single one, as typically only one or a few light curves would be available. Group normalization was applied, and then, in a series of successive convolution layers which decreased dimensionality, we reduced the data to 7 points across 256 channels.

Similar processing was independently done on the radial velocity, reducing the RV data from 512 points in a single channel, to 7 points in 256 channels, still keeping information on RVs of the primary and secondary separate.

The reduced data on the light curves and on RV curves were then concatenated, and further convolutional steps were applied, both horizontally and vertically, to produce a single point with 1024 channels.

Independently, broadband fluxes (converted from magnitudes to $\log \lambda F_\lambda$, sorted by wavelength, and normalized) were also subjected to convolution, effectively computing a combination of colors for a given star to understand the shape of the SED. Following all of the convolution steps, the result produced a single point with 512 channels. 

Afterwards, convolution outputs (from the light curves / RV curves and from broadband photometry) were then flattened, concatenated, also appending orbital period, parallax, and parallax uncertainty onto the resulting array.

This array was then passed through 5 fully connected layers with an output dimensionality of 1024. After each fully connected layer, the original array was concatenated onto the output. The final linear layer produced 42 outputs: 21 stellar and orbital parameters and their corresponding variances. The full architecture of this model is shown in Figure \ref{fig:model}.

The model was trained using the custom Gaussian negative log likelihood loss that would first create a normal distribution centered at the value of the predicted parameter,  with the corresponding predicted variance. The log probability was computed to evaluate the likelihood of the true labels given this distribution.

\subsection{Model 2 (PyTorch)}

Similar to Model 1, Model 2 handles the three input types with separate convolutional stacks, which are later combined, and final predictions are made with a fully connected feedforward network.

To aggregate across the 50 light curves, we applied a two-dimensional convolutional kernel spanning the entire light-curve axis to create 32 new channels. A max-pooling layer was then used to reduce the light-curve dimension to one. To improve convergence speed and stability, group normalization was applied to the resulting representation.

This was followed by a series of four convolutional blocks, each consisting of a 2-by-1 convolutional kernel, ReLU activation, and a 2-by-1 max-pooling layer. Each block doubled the number of channels, resulting in 512 channels after the four blocks. Group normalization was again applied to the resulting representation.

The radial velocities were processed in a similar way. A two-dimensional convolutional kernel was applied across the radial-velocity dimension to create 64 output channels, followed by a max-pooling layer that reduced the radial-velocity dimension to one. Group normalization was applied to the resulting representation. A series of four convolutional blocks with the same configuration as the light-curve branch was then applied, resulting in 1024 channels after the four blocks, with group normalization again applied.

The radial-velocity representation was reshaped to share a common channel width of 512 with the light-curve representation. The concatenated tensor was passed through a sequence of three two-dimensional convolutional blocks, configured similarly to those used in the light-curve and radial-velocity branches. The first and second blocks maintained the number of channels at 512, and the last block doubled the channels to 1024. Group normalization was again applied to the resulting representation.

The broadband fluxes were treated as a one-dimensional sequence. We applied a sequence of two convolutional blocks, each consisting of a convolutional layer, ReLU activation, max-pooling, and group normalization. The first convolutional block increased the number of channels from one to 128, and the second convolutional block maintained the same number of channels.

The flattened light-curve/radial-velocity representation, the flattened SED representation, the orbital period, parallax, and parallax uncertainty were concatenated to form the input to the final feedforward network. These features were projected to 2048 units, followed by a sigmoid activation, another hidden layer with sigmoid activation, and a final output layer producing the 42 outputs corresponding to 21 mean values and 21 variances.

The model was trained using Gaussian negative log-likelihood loss and the AdamW optimizer with a learning rate of 0.0001.

\section{Results}

\begin{figure*}
\epsscale{1.0}
\plotone{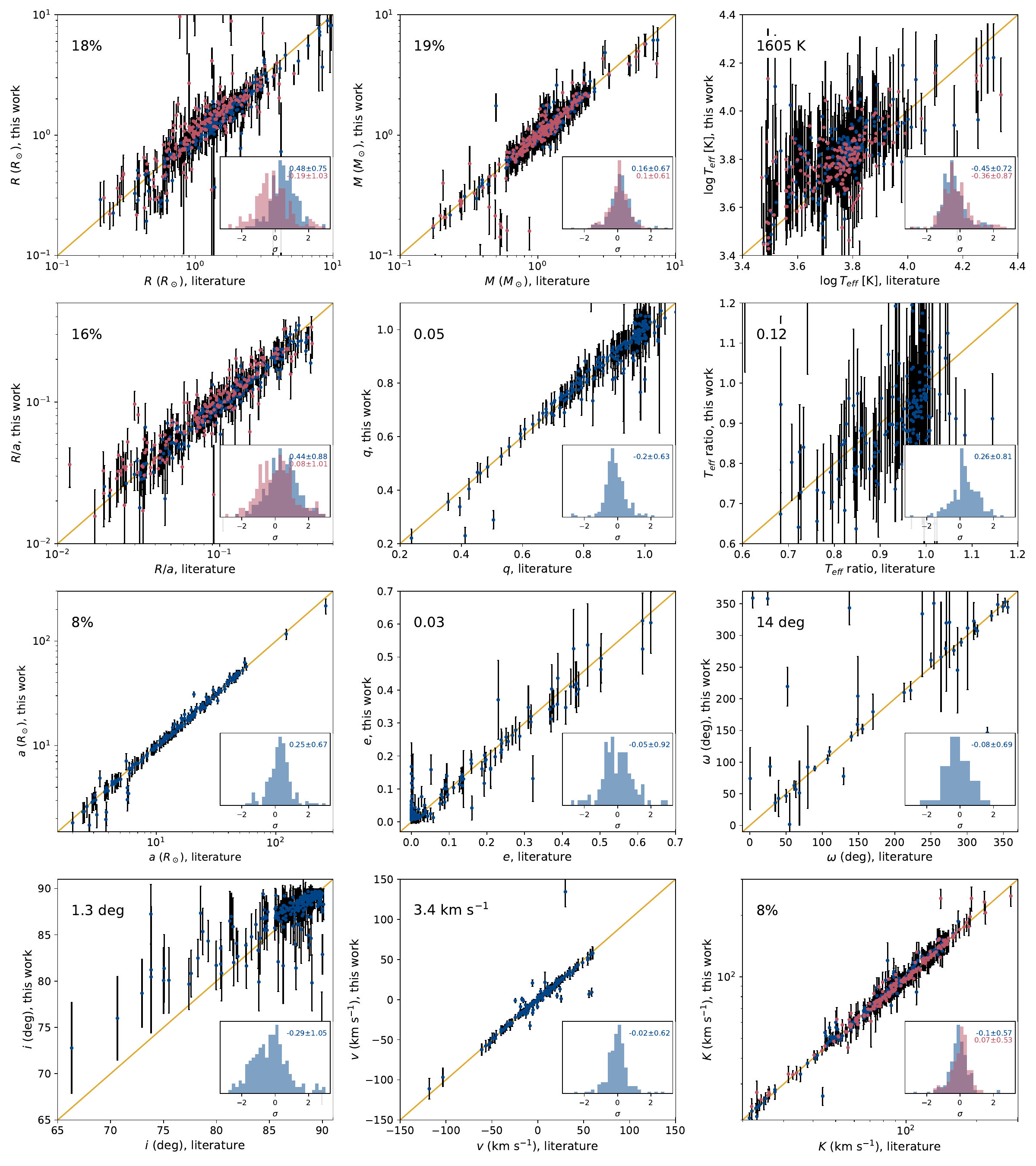}
\caption{Performance of the TensorFlow model on the real eclipsing binaries, showing a comparison between labels and the predictions for various parameters. Parameters of the primary star or the system as a whole are shown in blue, parameters of the secondary star are shown in red. The error bars show the uncertainties predicted by the model. The value in the top left corner of each panel shows the magnitude of the typical uncertainties in the sample. The inset histogram shows the characteristic scatter between the labels and the predictions, and the numbers in the top right corner indicate the mean and the standard deviation for the scatter. 
\label{fig:tensorflow}}
\end{figure*}

\begin{figure*}
\epsscale{1.0}
\plotone{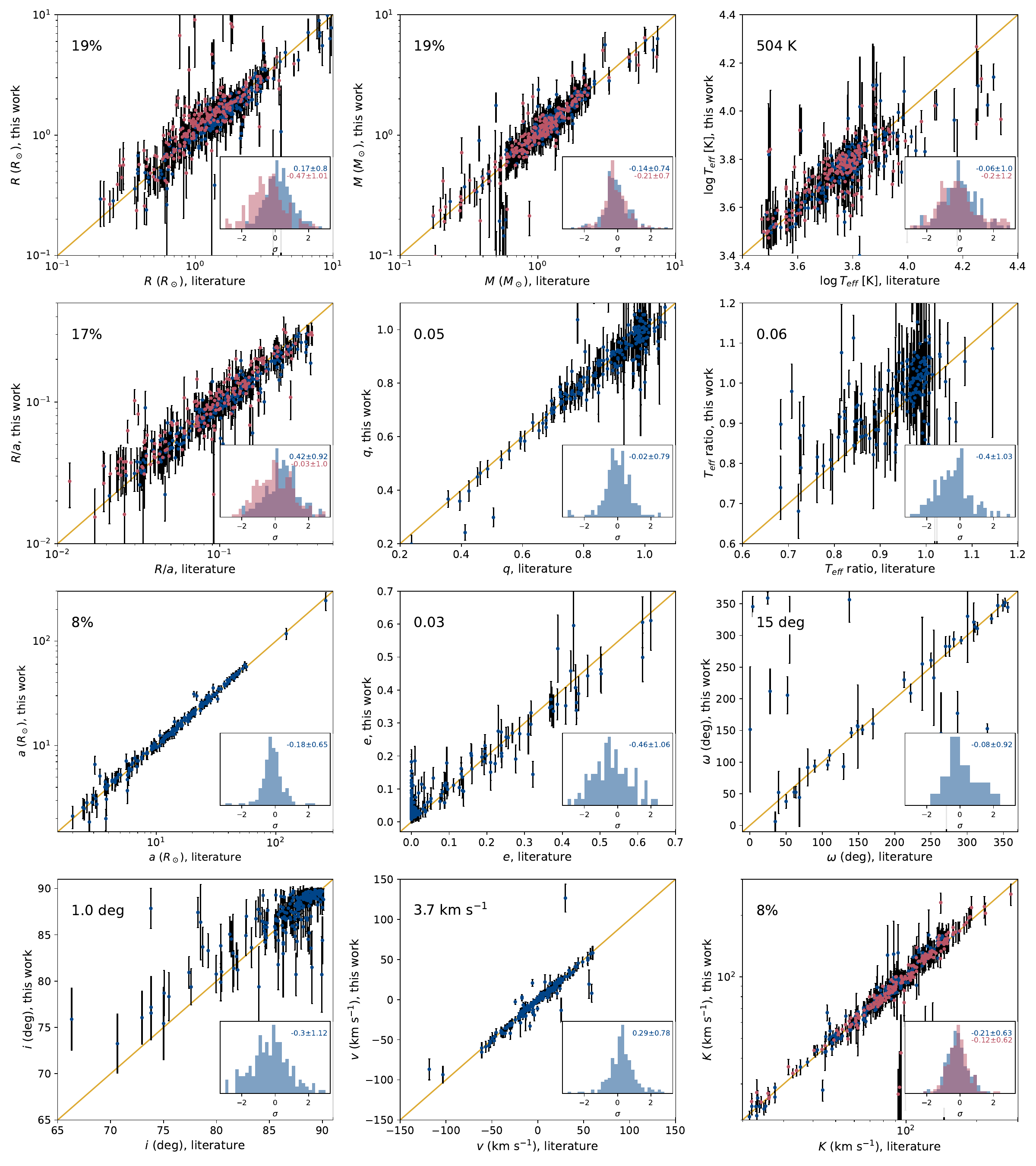}
\caption{Same as Figure \ref{fig:tensorflow}, but showing the performance of the PyTorch model.
\label{fig:pytorch}}
\end{figure*}

Both models have been trained on purely synthetically generated data, but to test their efficacy we apply them to the real EBs described in Section \ref{sec:real}. Figures \ref{fig:tensorflow} and \ref{fig:pytorch} show the comparison of the resulting predictions (alongside with generated uncertainties) relative to the parameters derived in the literature for these systems. Figures \ref{fig:tensorflow_norv} and \ref{fig:pytorch_norv} show this comparison, but all radial velocities were censored, and evaluations were done solely on the light curves and broadband photometry -- i.e., the data which are most likely to be available for most EBs. The estimated uncertainties are a reflection of the variance that exists in the synthetic data and our noise generating process. We believe that the distribution of our training data is sufficiently similar to the real data to produce uncertainties comparable to those reported in the literature. Unlike approaches that estimate the posterior over stellar parameters, such as MCMC, the uncertainties we report do not reflect uncertainty in our model parameters.

\subsection{Uncertainties}
Overall, there is a strong agreement between literature values (labels) and the predictions, all strongly follow the 1-1 trend line. Moreover, the uncertainties appear to reliably reproduce the underlying scatter between the labels and predictions: the typical scatter is 0.7$\sigma$ for the TensorFlow model, and 0.85$\sigma$ for the PyTorch model. The typical scatter exceeds 1 $\sigma$ in only a few  parameters. This suggests that uncertainties may be modestly overestimated, in some cases by as much as a factor of 2.

The predicted uncertainties are generally larger than the quoted uncertainties in the literature. E.g., DEBCat \citep{southworth2015} only curates the sources that have been reported to have uncertainties in mass and in radius of $<2$\%, while our models produce typical uncertainties of $\sim$20\%. Even if we account for the overestimation of uncertainties relative to scatter, it would only reduce them to $\sim$13\%.

There are several factors that likely influence this disparity. First of all, in most cases, we are not using the same underlying dataset, frequently having to independently source light curves, RVs, or both, and thus the scatter and the appropriate uncertainties for the respective datasets may be different. Second, because our model is designed to handle almost all EBs (including EA, EB, and EW, with all separations, eccentricities, and mass ratios), higher scatter relative to the ground truth may be expected than with a more boutique analysis of an individual source. And third, in many cases in the literature (particularly as a considerable number of papers with stellar parameters date back to the 1980s and 1990s), uncertainties may be somewhat underestimated relative to the true expected scatter, not taking into the account more systematic errors. Thus, although a more careful ``solving'' of the systems using traditional techniques may improve on the predictions (and having a robust initial estimate should streamline the process of performing such a modeling), there is likely a floor beyond which, realistically, the uncertainties may not improve further.

It is worth noting that our DNN only models aleotoric uncertainty (i.e., instrinsic to the data), and not epistemic uncertainty resulting from the model estimation. As a result, our uncertainty estimates, while high, may be {\it lower} than they would be had we explicitly modeled uncertainty in our DNN parameter estimates, or uncertainty arising from a domain shift between the training and test data.

\begin{figure*}
\epsscale{1.0}
\plotone{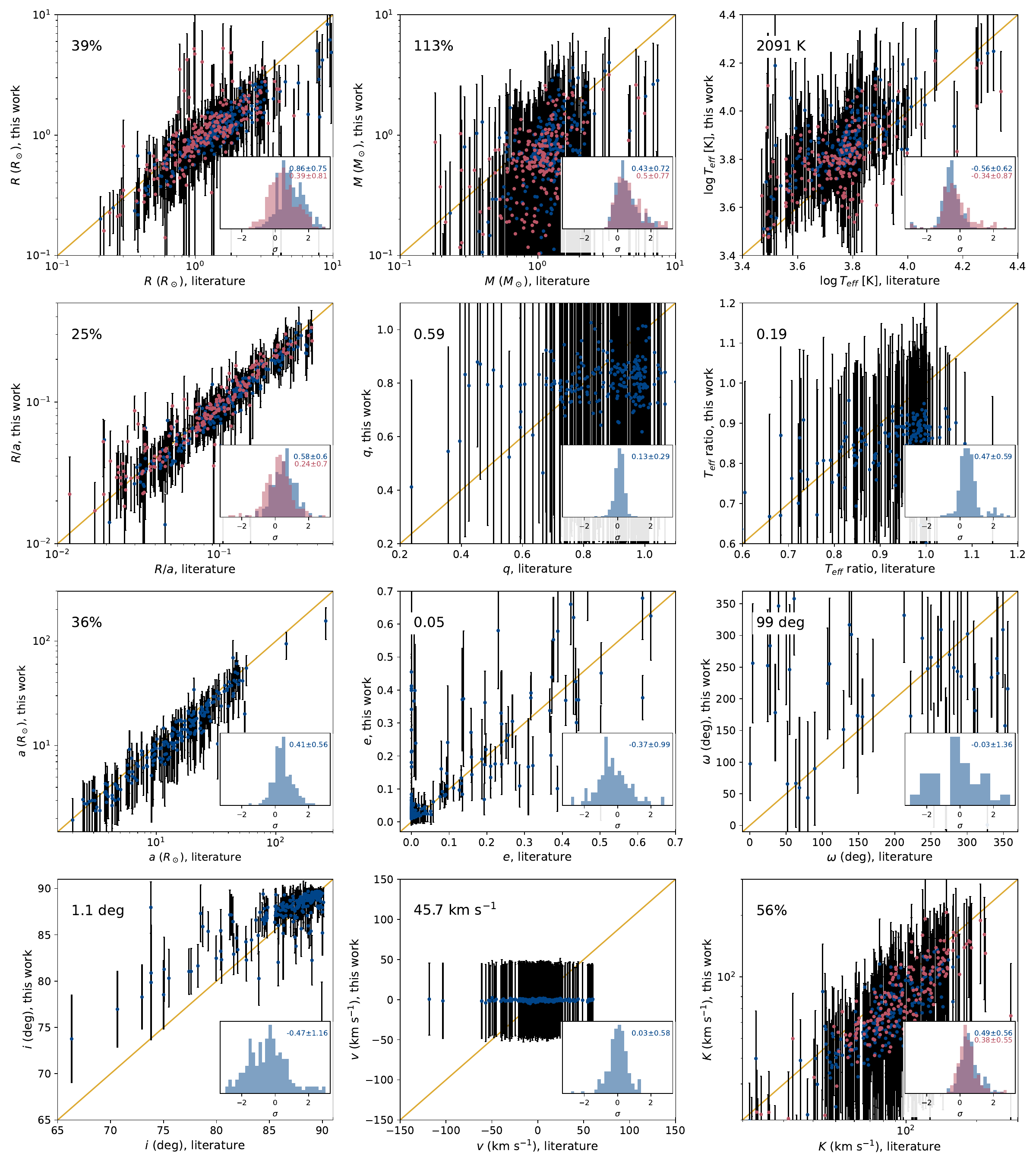}
\caption{Same as Figure \ref{fig:tensorflow}, showing TensorFlow model, but all of the radial velocities were masked in the data, thus the predictions were done from just the light curves and SED.
\label{fig:tensorflow_norv}}
\end{figure*}

\begin{figure*}
\epsscale{1.0}
\plotone{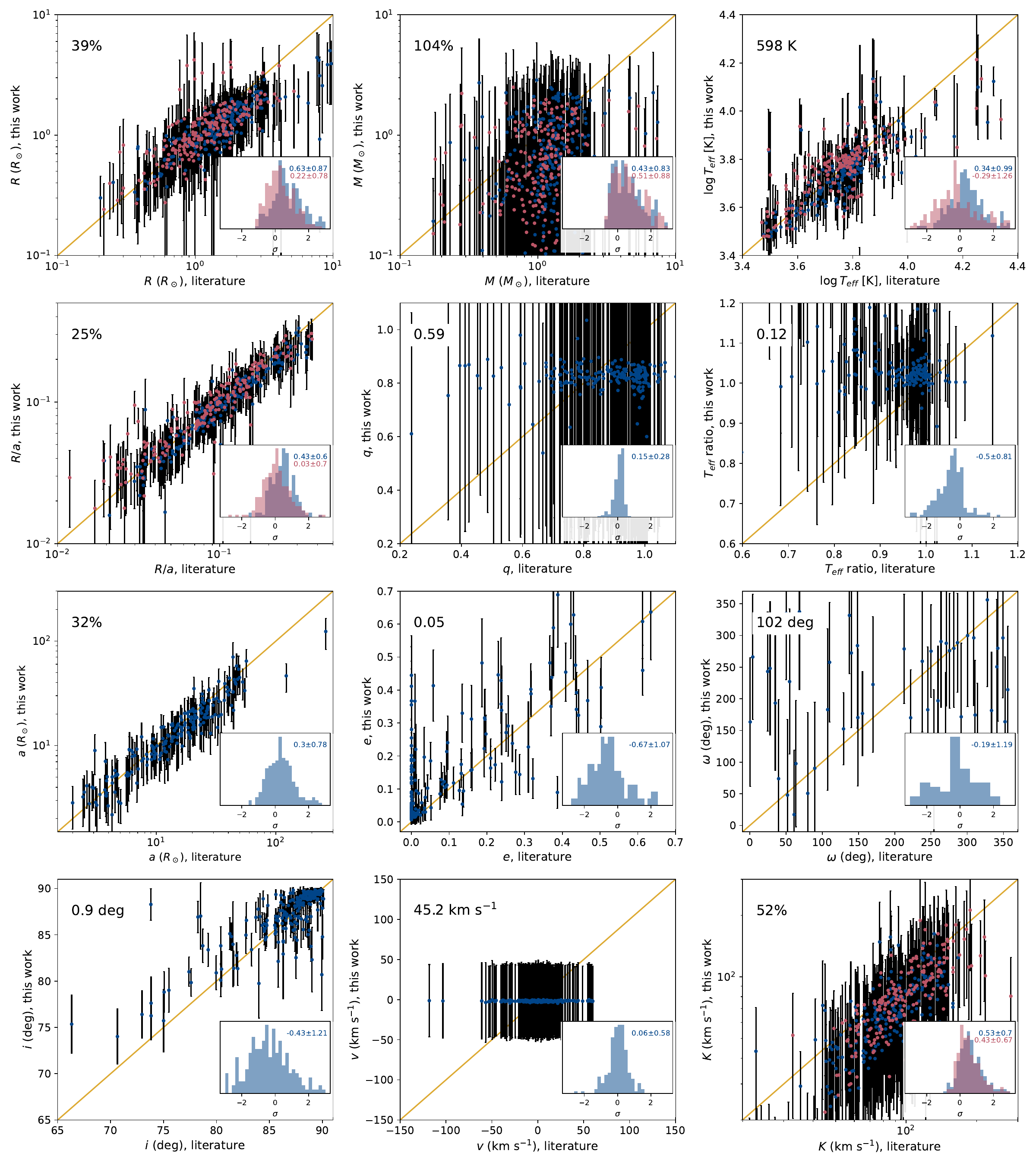}
\caption{Same as Figure \ref{fig:tensorflow}, showing PyTorch model, but all of the radial velocities were masked in the data, thus the predictions were done from just the light curves and SED.
\label{fig:pytorch_norv}}
\end{figure*}

Reliability of predicted uncertainties relative to the predicted values is particularly pronounced in the evaluations without RVs. Without RVs (Figures \ref{fig:tensorflow_norv} and \ref{fig:pytorch_norv}), it is difficult to infer $q$ in any manner other than statistical (particularly as the model was deliberately prevented from relating masses and \teff\ in the manner the training set was constructed). This is unsurprising, as no approach is able to infer $q$ just from the light curves, information on kinematics is needed to obtain it. With poorer $q$, masses and velocity amplitudes become more uncertain. And, although the ratio of radii to semimajor axis can still be inferred as well as in the case with RVs, both the semimajor axis itself, as well as radii also become more uncertain. However, all predictions stay reasonably close to the 1-1 line, as there is only a finite range of $q$ that a system detected as EB would be likely to have.

The most demonstrative case of the model faithfully reproducing the uncertainties is $\gamma$. Without any RVs, the model is unable to generate any type of accurate predictions, consistently returning values of $\gamma\sim0\pm45$ \kms\ --- well-encompassing the distribution of the possible RVS for the field stars. 

Eccentricity is another parameter that suffers without RVs, as they respond more to the more eccentric orbits. With just the light curves, it is straightforward to find $e\cos\omega$ from the separation between the eclipses, but $e\sin\omega$ is only seen through a slight asymmetry between the shapes of the primary and secondary eclipses, which may be too subtle to detect reliably. Although most cases still show a good agreement, there are a handful of non-eccentric sources which get mistaken for having non-zero eccentricity, requiring additional vetting.

Curiously, we find \teff\ ratio to be poorly constrained without RVs. It is traditionally seen as a parameter that can be inferred just from the ratio of the eclipse depths \citet{prsa2008}, corresponding to the ratio of fluxes. However, the ratio of fluxes corresponds to both the ratio of \teff s, and the ratio of radii. More uncertain radii make it difficult to directly infer \teff\ ratio.

To better demonstrate how RVs influence the degree of scatter in the parameters, we also perform a test where we drop most of the RVs with an exception of a set number of data points for a given system. The comparison is done relative to the predictions with the full complement of data. Unsurprisingly, the scatter is largest with the fewest number of RV measurements, and it steadily decreases as additional RV measurements are added.

Similarly to withholding RVs, we have also done a test of withholding broadband fluxes of the SED. The only parameters that suffer meaningfully without them are \teff. While \teff\ ratio can still be recovered, \teff\ of individual sources have typical uncertainties of $\sim$5000 K. This is consistent with more traditional techniques of deriving stellar parameters from EBs, \teff\ is poorly constrained from the light curves alone \citep[e.g.,][]{kounkel2024a}.

We note, however, that while the performance of EBNet degrades if RVs or fluxes are not available, this issue is not unique to EBNet, it is intrinsic to any analysis that involves trying to solve EBs using just the light curves, leading these analyses to curate a more limited set of parameters that could be inferred solely from the light curves. Because EBNet does report uncertainties which qualitatively change with missing data, it is able to demonstrate more directly the exact range of useful parameters for a given system.

\subsection{Comparison between TensorFlow and PyTorch models} \label{sec:comparison}

Both PyTorch and TensorFlow models do have a comparable performance, but, nonetheless, there are particular parameters where a particular model has a superior performance.

In particular, the \teff\ and \teff\ ratio are significantly more reliable in the PyTorch model, achieving a typical precision of $\sim$500 K, in contrast to $\sim$1650 K in TensorFlow. Furthermore, the 1-1 plot shows fewer systematic differences between the labels and the predictions in PyTorch, while TensorFlow appears to systematically overestimate \teff\ at 4000 K. Additionally, PyTorch achieves a somewhat improved performance for the semimajor axis, and for the inclination. It is not clear what specifically about the model implementation that produces this difference, as it has persisted even in tests with an identical model architecture. 

On the other hand, for other parameters, PyTorch is somewhat more strongly affected by the noisy data producing more outliers, as well as having larger scatter compared to the labels, with larger uncertainties, and TensorFlow predictions are preferred for the remaining parameters.

Since we have been unable to come up with a single model that produces a superior performance across the board, instead we ended up combining both models into one. To make it possible to use only a single code base, the weights of the model that was trained in TensorFlow was converted into PyTorch using Onnx. The translated model has identical outputs as the original when applied to the same data.

During execution, EBNet evaluates both the original PyTorch and the translated TensorFlow models by default, and selects the best model for each parameter. As all of the parameters are evaluated completely independently of one another, this mixing and matching does not add any particular biases in the predictions.

\section{Conclusions}

We developed a new neural network, EBNet, for ``solving'' eclipsing binary systems to obtain their stellar and orbital parameters. Compared to existing tools, it is fast (taking $<$1 second to make its predictions, unlike the full MCMC in solvers such as PHOEBE), and it is capable of simultaneously handling multiple passbands, radial velocities, as well as other additional metadata such as fluxes across the SED (unlike the estimators such as EBAI). The model has been trained to be agnostic to the presence of spots as well as the third light, both having been injected into the training sample.

More importantly, EBNet is capable of producing uncertainty estimates that are in strong agreement with those reported in the literature for all parameters predicted for each individual source. These uncertainties may be modestly overestimated, but they scale with the quality of the data, and they also are able to handle missing data - as such even with completely absent RV information it is possible to produce reasonably accurate estimates for most parameters, such as radii (and even to some degree, velocity amplitude for the individual stars) with appropriate uncertainties, just from the light curve.

However, we find that having SED is absolutely crucial to reliably determining \teff. Even when multiple light curves are available in different passbands, SED is much more effective for measuring the underlying \teff\ for individual stars; this is the case not only in EBNet, but also for more traditional techniques as well.

Although EBNet has been trained on purely synthetic data, we have tested it on a set of $\sim200$ real eclipsing binaries with known stellar and orbital parameters, and overall it appears to be able to successfully recover these parameters. Although the predictions of EBNet are somewhat more uncertain than the parameters evaluated in the boutique traditional analysis that is tailored to each individual star, the speed and reliability of our method is better suited for the explosion of data of EBs from large photometric monitoring surveys, such as Kepler, TESS, ZTF, LSST, and others, to at least help with determining parameters for most systems, to identify systems of interest for follow-up, whether it is to select systems for which a full fit should be performed, or to select systems for follow-up spectroscopic observations.

EBNet also makes it possible to calibrate the possible distributions for the initial guesses for systems that would benefit from the more traditional analysis. That is to say, in fully solving an eclipsing binary system, PHOEBE \citep{conroy2020} recommends to first use an estimator before attempting to use a sampler, and EBNet significantly improves over the currently available estimators.

\software{TOPCAT \citep{topcat}, TensorFlow, PyTorch, EBNet \citep{ebnet}}

\acknowledgments

MK acknowledges support provided by NASA grant 80NSSC24K0620.

This work has made use of data from the European Space Agency (ESA)
mission {\it Gaia} (\url{https://www.cosmos.esa.int/gaia}), processed by
the {\it Gaia} Data Processing and Analysis Consortium (DPAC,
\url{https://www.cosmos.esa.int/web/gaia/dpac/consortium}). Funding
for the DPAC has been provided by national institutions, in particular
the institutions participating in the {\it Gaia} Multilateral Agreement.

\bibliographystyle{aasjournal.bst}
\bibliography{main.bbl}

\begin{thebibliography}{}
\expandafter\ifx\csname natexlab\endcsname\relax\def\natexlab#1{#1}\fi
\providecommand{\url}[1]{\href{#1}{#1}}
\providecommand{\dodoi}[1]{doi:~\href{http://doi.org/#1}{\nolinkurl{#1}}}
\providecommand{\doeprint}[1]{\href{http://ascl.net/#1}{\nolinkurl{http://ascl.net/#1}}}
\providecommand{\doarXiv}[1]{\href{https://arxiv.org/abs/#1}{\nolinkurl{https://arxiv.org/abs/#1}}}

\bibitem[{{Conroy} {et~al.}(2020){Conroy}, {Kochoska}, {Hey}, {Pablo},
  {Hambleton}, {Jones}, {Giammarco}, {Abdul-Masih}, \&
  {Pr{\v{s}}a}}]{conroy2020}
{Conroy}, K.~E., {Kochoska}, A., {Hey}, D., {et~al.} 2020, \apjs, 250, 34,
  \dodoi{10.3847/1538-4365/abb4e2}

\bibitem[{{Eker} {et~al.}(2014){Eker}, {Bilir}, {Soydugan}, {G{\"o}k{\c{c}}e},
  {Soydugan}, {T{\"u}ys{\"u}z}, {{\c{S}}eny{\"u}z}, \& {Demircan}}]{eker2014}
{Eker}, Z., {Bilir}, S., {Soydugan}, F., {et~al.} 2014, \pasa, 31, e024,
  \dodoi{10.1017/pasa.2014.17}

\bibitem[{Gordon(2024)}]{dustextinction}
Gordon, K. 2024, dust\_extinction, v1.4.1,  Zenodo,
  \dodoi{10.5281/zenodo.11235336}

\bibitem[{{Husser} {et~al.}(2013){Husser}, {Wende-von Berg}, {Dreizler},
  {Homeier}, {Reiners}, {Barman}, \& {Hauschildt}}]{husser2013}
{Husser}, T.-O., {Wende-von Berg}, S., {Dreizler}, S., {et~al.} 2013, \aap,
  553, A6, \dodoi{10.1051/0004-6361/201219058}

\bibitem[{{Jenkins} {et~al.}(2016){Jenkins}, {Twicken}, {McCauliff},
  {Campbell}, {Sanderfer}, {Lung}, {Mansouri-Samani}, {Girouard}, {Tenenbaum},
  {Klaus}, {Smith}, {Caldwell}, {Chacon}, {Henze}, {Heiges}, {Latham},
  {Morgan}, {Swade}, {Rinehart}, \& {Vanderspek}}]{jenkins2016}
{Jenkins}, J.~M., {Twicken}, J.~D., {McCauliff}, S., {et~al.} 2016, in Society
  of Photo-Optical Instrumentation Engineers (SPIE) Conference Series, Vol.
  9913, Software and Cyberinfrastructure for Astronomy IV, ed. G.~{Chiozzi} \&
  J.~C. {Guzman}, 99133E, \dodoi{10.1117/12.2233418}

\bibitem[{{Kounkel}(2023)}]{sedfit}
{Kounkel}, M. 2023, SEDFit,  Zenodo, \dodoi{10.5281/zenodo.8076500}

\bibitem[{{Kounkel} \& {Stassun}(2024)}]{kounkel2024a}
{Kounkel}, M., \& {Stassun}, K.~G. 2024, \aj, 168, 134,
  \dodoi{10.3847/1538-3881/ad6a17}

\bibitem[{{Kounkel} {et~al.}(2024){Kounkel}, {Statti}, {Kulkarni}, {Stassun},
  \& {Sun}}]{kounkel2024}
{Kounkel}, M., {Statti}, M., {Kulkarni}, A., {Stassun}, K.~G., \& {Sun}, M.
  2024, \mnras, 527, 3806, \dodoi{10.1093/mnras/stad3439}

\bibitem[{{Lightkurve Collaboration} {et~al.}(2018){Lightkurve Collaboration},
  {Cardoso}, {Hedges}, {Gully-Santiago}, {Saunders}, {Cody}, {Barclay}, {Hall},
  {Sagear}, {Turtelboom}, {Zhang}, {Tzanidakis}, {Mighell}, {Coughlin}, {Bell},
  {Berta-Thompson}, {Williams}, {Dotson}, \& {Barentsen}}]{lightkurve}
{Lightkurve Collaboration}, {Cardoso}, J.~V.~d.~M., {Hedges}, C., {et~al.}
  2018, {Lightkurve: Kepler and TESS time series analysis in Python},
  Astrophysics Source Code Library.
\newblock \doeprint{1812.013}

\bibitem[{{Parviainen} \& {Aigrain}(2015)}]{parviainen2015}
{Parviainen}, H., \& {Aigrain}, S. 2015, \mnras, 453, 3821,
  \dodoi{10.1093/mnras/stv1857}

\bibitem[{{Pr{\v{s}}a} {et~al.}(2008){Pr{\v{s}}a}, {Guinan}, {Devinney},
  {DeGeorge}, {Bradstreet}, {Giammarco}, {Alcock}, \& {Engle}}]{prsa2008}
{Pr{\v{s}}a}, A., {Guinan}, E.~F., {Devinney}, E.~J., {et~al.} 2008, \apj, 687,
  542, \dodoi{10.1086/591783}

\bibitem[{{Pr{\v{s}}a} \& {Zwitter}(2005)}]{prsa2005}
{Pr{\v{s}}a}, A., \& {Zwitter}, T. 2005, \apj, 628, 426, \dodoi{10.1086/430591}

\bibitem[{Sizemore \& Kounkel(2026)}]{ebnet}
Sizemore, L., \& Kounkel, M. 2026, hutchresearch/EBNet, v1.0,  Zenodo,
  \dodoi{10.5281/zenodo.18790155}

\bibitem[{{Southworth}(2015)}]{southworth2015}
{Southworth}, J. 2015, in Astronomical Society of the Pacific Conference
  Series, Vol. 496, Living Together: Planets, Host Stars and Binaries, ed.
  S.~M. {Rucinski}, G.~{Torres}, \& M.~{Zejda}, 164.
\newblock \doarXiv{1411.1219}

\bibitem[{{Taylor}(2005)}]{topcat}
{Taylor}, M.~B. 2005, in Astronomical Society of the Pacific Conference Series,
  Vol. 347, Astronomical Data Analysis Software and Systems XIV, ed.
  P.~{Shopbell}, M.~{Britton}, \& R.~{Ebert}, 29

\bibitem[{{Wrona} \& {Pr{\v{s}}a}(2025)}]{wrona2025}
{Wrona}, M., \& {Pr{\v{s}}a}, A. 2025, \apjs, 277, 1,
  \dodoi{10.3847/1538-4365/ada4ae}

\end{thebibliography}

\end{document}